%% file: Digital_Signal_Processing.tex
\documentclass[preview]{elsarticle}

\usepackage{hyperref}
{}
{}
{}

\usepackage{amssymb}
\usepackage{amsmath}
\usepackage{upgreek}
\usepackage[section]{placeins}
\usepackage{graphicx}
\usepackage{lineno,hyperref}
\usepackage{makecell}
\usepackage{algorithm}
\usepackage{algorithmic}
\usepackage{booktabs}
\usepackage{tabularx}
\usepackage{bbding}
\usepackage{amsthm}
\usepackage[misc]{ifsym}
\usepackage{array}
\usepackage{lineno}
\usepackage{color,soul}
\soulregister\ref{7}
\soulregister\eqref{7}
\soulregister\cite{7}
\soulregister\mathbf{7}

\begin{document}
	
	\begin{frontmatter}
		
		\title{A Sequential MUSIC algorithm for Scatterers Detection in SAR Tomography Enhanced by a Robust Covariance Estimator}
		
		\author[mymainaddress]{Ahmad Naghavi\corref{mycorrespondingauthor}}
		\cortext[mycorrespondingauthor]{Corresponding author}
		\ead{a.naghavi@ec.iut.ac.ir}
		
		\author[mymainaddress]{Mohammad Sadegh Fazel}
		\author[mysecondaryaddress]{Mojtaba Beheshti}
		\author[mymainaddress]{Ehsan Yazdian}

		\address[mymainaddress]{Department of Electrical and Computer Engineering, Isfahan University of Technology, Isfahan, Iran}
		\address[mysecondaryaddress]{Information and Communication Technology Institute, Isfahan University of Technology, Isfahan, Iran}

		\begin{abstract}
			Synthetic aperture radar (SAR) tomography (TomoSAR) is an appealing tool for the extraction of height information of urban infrastructures. Due to the widespread applications of the MUSIC algorithm in source localization, it is a suitable solution in TomoSAR when multiple snapshots (looks) are available. While the classical MUSIC algorithm aims to estimate the whole reflectivity profile of scatterers, sequential MUSIC algorithms are suited for the detection of sparse point-like scatterers. In this class of methods, successive cancellation is performed through orthogonal complement projections on the MUSIC power spectrum. In this work, a new sequential MUSIC algorithm named recursive covariance canceled MUSIC (RCC-MUSIC), is proposed. This method brings higher accuracy in comparison with the previous sequential methods at the cost of a negligible increase in computational cost. Furthermore, to improve the performance of RCC-MUSIC, it is combined with the recent method of covariance matrix estimation called correlation subspace. Utilizing the correlation subspace method results in a denoised covariance matrix which in turn, increases the accuracy of subspace-based methods. Several numerical examples are presented to compare the performance of the proposed method with the relevant state-of-the-art methods. As a subspace method, simulation results demonstrate the efficiency of the proposed method in terms of estimation accuracy and computational load.

		\end{abstract}
		
		
	\end{frontmatter}	
	
	\section{Introduction}\label{sec1}
	Synthetic aperture radar (SAR) facilitates a day-night, all-weather, two-dimensional imaging for earth observation \cite{moreira2013tutorial}. Using three-dimensional (3-D) SAR imaging, the distribution of scatterers in the elevation direction is reached. SAR tomography (TomoSAR) is capable of resolving the scatterers through a synthetic aperture in the elevation direction made by multi-temporal SAR acquisitions\cite{reigber2000first,fornaro2005three}. This technique, which was primarily used for obtaining information of volumetric environments, is widely applied for urban infrastructure monitoring where the scatterers are sparsely distributed\cite{fornaro2014tomographic}. By using tomography as a complement to interferometric techniques like persistent scatterer interferometry (PSI), the detection ability of the point-like scatterers improves \cite{siddique2016single}.
	
	The tomographic information are extracted using a spectral estimation method such as conventional beamforming \cite{reigber2000first}, TSVD \cite{fornaro2003three}, Capon \cite{lombardini2003adaptive} and MUSIC \cite{guillaso2005polarimetric}. For urban environments, the scatterers are discrete point-like, hence the estimation of the profile of scatterers reduces to the problem of detection of some sparse scatterers (targets) \cite{budillon2016glrt}. SGLRTC \cite{pauciullo2012detection},  NLS \cite{zhu2010very} and SL1MMER \cite{zhu2012super} have been proposed for multiple scatters detection in TomoSAR. These methods which are used in the single snapshot scenario have also been extended for the case of multiple snapshots of data \cite{pauciullo2018multi}. 
	
	The MUSIC method as a solution for multiple snapshots scenario has a high-resolution capability with moderate computational cost \cite{schmide1979multiple}. MUSIC is widely used for DoA estimation in array signal processing \cite{stoica1989music,krim1996two} where it is evolved by numerous modified algorithms to enhance both complexity and the accuracy \cite{kundu1996modified,zhang2018localization,asghari2022ecf}. It has been used for scatterers profile retrieval in TomoSAR applications \cite{guillaso2005polarimetric,lombardini2003reflectivity}. For the application of sparse support recovery, the MUSIC algorithm is used in a greedy manner similar to the orthogonal matching pursuit (OMP) algorithm \cite{tropp2006algorithms,dai2009subspace}. This class, called sequential MUSIC or recursive MUSIC methods, facilitates the detection of point targets through a successive cancellation routine. Several sequential music algorithms have been presented in the literature, among them are sequential MUSIC (S-MUSIC) \cite{oh1993sequential}, improved sequential MUSIC (IES-MUSIC) \cite{stoica1995improved} and recursively applied and projected MUSIC (RAP-MUSIC) \cite{mosher1999source}. The main procedure in all of these methods is successive cancellation of the detected targets through maximizing the projection onto the signal subspace. Rank Aware Order Recursive Matching Pursuit (RA-ORMP) is another MUSIC-based algorithm that is algorithmically similar to RAP-MUSIC but is presented for sparse support recovery applications \cite{davies2012rank}. It is shown that RA-ORMP is superior to simultaneous OMP (SOMP) which is the multi-snapshot variant of OMP \cite{davies2012rank}. As the resolution of greedy-based methods is limited, some methods are proposed for resolution enhancement by extra processing on the initial resulted locality information \cite{dehghani2018fomp,naghavi2020super,zhao2021direction}.

	The performance of MUSIC as a subspace-based method highly depends on the estimation accuracy of the covariance matrix. Sample covariance matrix (SCM) is the conventional method for estimation of covariance matrix which is the asymptotic maximum likelihood (ML) estimate where the number of samples tends to infinity. For a small number of snapshots, the resulting signal and noise subspaces by SCM deviate from the real subspace., i.e the subspace leakage appears between the signal and noise subspaces \cite{shaghaghi2015subspace,yazdian2012source}. Consequently, in the small sample size regime, the estimation bias of SCM deteriorates the performance of subspace-based methods. Various methods have been proposed for a consistent estimation of the covariance matrix. The covariance matrix is usually estimated by imposing either Toeplitzness \cite{wikes1988iterated,li1999computationally,chachlakis2021structured,wang2021fda}, low rank \cite{lounici2014high,chu2003structured,liu2021rank,hippert2022robust} and sparseness \cite{bien2011sparse,romero2015compressive,prasanna2021mmwave} constraints on the solution. Also, positive semi-definiteness is a necessary property of covariance matrix which is used in most of the covariance estimation methods \cite{burg1982estimation,jansson2000structured}.
	
	Recently a high-quality covariance matrix estimation approach named correlation subspace is proposed \cite{rahmani2016subspace}. In this approach, first, a solution is given for finding a subspace that spans the vectorized denoised covariance matrix. Then by imposing some constraints using the estimated subspace on a least-square fitting problem, a denoised version of the covariance matrix is reached. It has been shown that applying this covariance matrix estimation improves the quality of subspace-based methods like MUSIC and Capon \cite{rahmani2016subspace}. The correlation subspace method has been implemented in two ways, optimal and suboptimal \cite{rahmani2016subspace,liu2017correlation}. While the optimal solution is computationally complex due to the need for optimizing a convex problem, the suboptimal one gives satisfactory results which only requires a further matrix multiplication.
	
	In this paper, we investigate the estimation accuracy of the location of two scatterers in a range-azimuth cell using a sequential MUSIC algorithm. To this end, first, we apply the correlation subspace method for the estimation of the covariance matrix. Then we propose a sequential MUSIC algorithm called the RCC-MUSIC algorithm which will be shown to have higher accuracy compared to the other sequential MUSIC algorithms. Deploying the denoised covariance matrix estimate in the RCC-MUSIC method, the targets are more detectable with lower error values. In this paper, the number of scatterers is estimated by information-theoretic criteria (ITC) \cite{wax1985detection}. The performance of the methods is evaluated in terms of estimation accuracy and speed of implementation. The combination of the proposed sequential MUSIC algorithm together with the denoising covariance matrix estimator enables resolving closely-spaced targets at a lower cost compared to ML and $l_1$-minimization methods. Although our method is proposed for the TomoSAR application, it can be deployed for any general spectral estimation application in multiple snapshots scenario.
	
	The rest of this paper is organized as follows. Section \ref{sec2} introduces the system model and states the problem formulation. Section \ref{sec3} reviews the MUSIC algorithm and describes the correlation subspace method for covariance matrix estimation. Section \ref{sec4} introduces the proposed method that is in the category of sequential MUSIC algorithms. Section \ref{sec5} investigates the performance of the proposed method and assesses the impact of the applied covariance matrix estimator on the estimation accuracy. Finally, Section \ref{sec6} concludes this work.
	\begin{figure}
		\centering
		\includegraphics[width=0.70\linewidth]{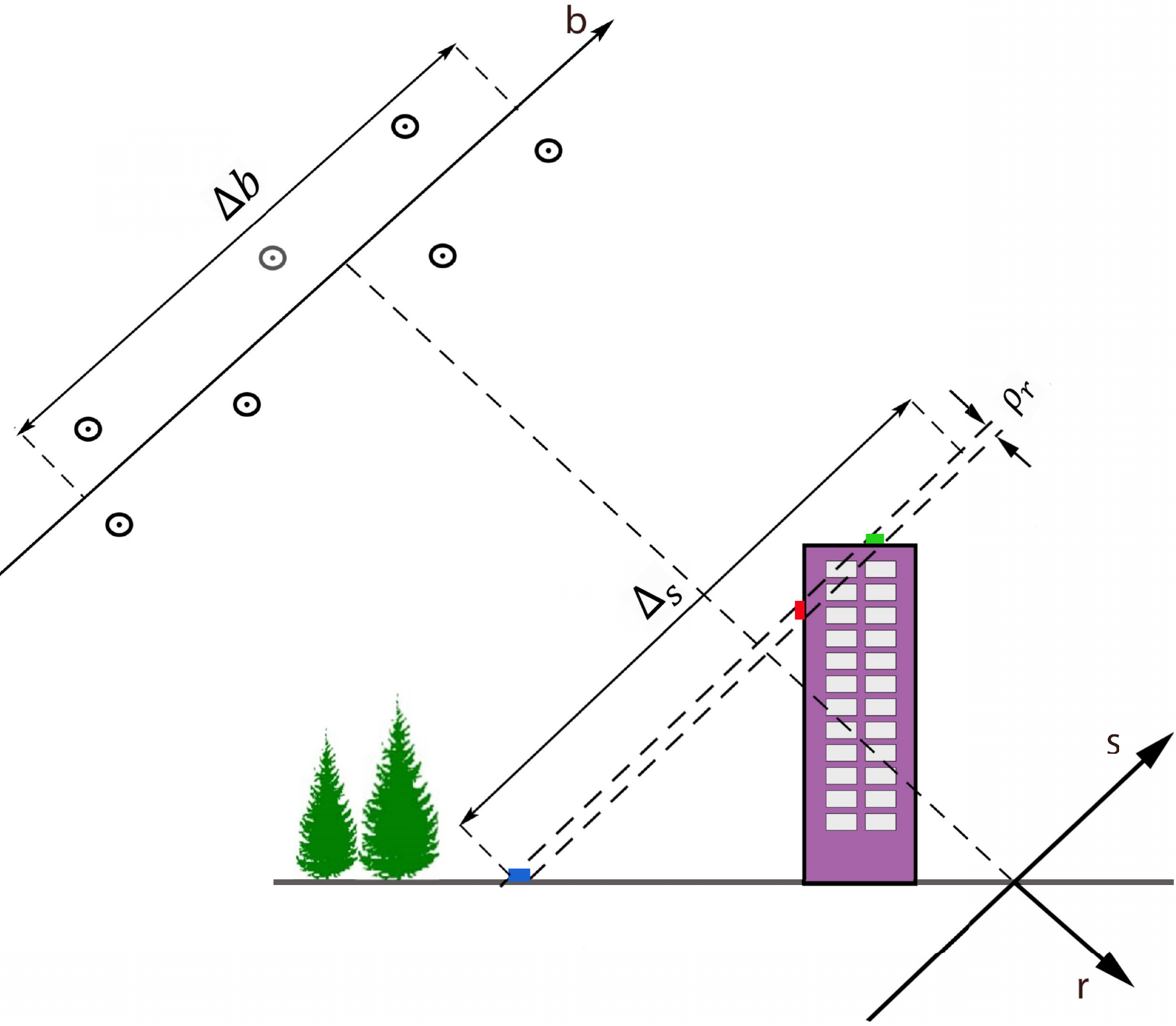}
		\caption{Multipass SAR geometry in the range-elevation $(r,s)$ plane.}
		\label{fig:tomosar-geometry}
	\end{figure}
	\section{System model and problem formulation}\label{sec2}
	The multipass SAR geometry is shown in Fig. \ref{fig:tomosar-geometry}. TomoSAR processing starts with acquiring a stack of $N$ range-azimuth compressed complex images from slightly different orbits. After aligning the image stack with respect to a reference -master- orbit and applying some postprocessing (deramping, phase calibration,...), a complex-valued vector is prepared for every range-azimuth pixel. Assuming $k$ point-like scatterers in the elevation coordinate $s$,  it is shown that the following relationship is held for every considered pixel,  \cite{fornaro2005three}
	\begin{equation}
		\label{Eq_1}
		\mathbf{g}(l) = \sum_{i=1}^{k}\gamma_{m_i}(l) \boldsymbol{a}(s_{m_i}) + \pmb{\it{w}}(l), \hspace{1.03cm} l=1,2,\ldots,L,
	\end{equation}
	where $\mathbf{g}(l)$ is the $N\times{1}$ vector of the range-azimuth compressed images, $m_i$ is the location of $i$-th scatterer in the elevation and $\gamma_{m_i}$ is the reflectivity of the related scatterer where $m_i \in [1,M]$, as $M$ is number of grids in elevation. $\boldsymbol{a}(s_m)$ is the $N\times{1}$ steering vector whose $n$-th element is given by
	\begin{equation}
		\label{Eq_2}
		[\boldsymbol{a}(s_m)]_n=\frac{1}{N}\exp(j2\pi\xi_{n}s_{m}),
	\end{equation}	
	where $\xi_n$ is the spatial frequency and $j=\sqrt{-1}$. Accordingly, the steering matrix is defined as  $\mathbf{A}=[\boldsymbol{a}(s_1),\boldsymbol{a}(s_2),... ,\boldsymbol{a}(s_M)]$.
	
	The noise term $\pmb{\it{w}}(l)$ is a zero-mean circular complex Gaussian vector with covariance matrix $E{\{\pmb{\it{w}}\pmb{\it{w}}^\mathrm{H}}\}=\sigma^2_n\mathbf{I}_N$, where $\mathbf{I}_N$ is $N\times{N}$ identity matrix. The temporal argument $(l)$ denotes the $l$-th snapshot as a multi-snapshot acquisition mode where $L$ is the number of snapshots or equivalently, number of looks in TomoSAR.

	The basic tomographic estimation method is conventional beamforming \cite{reigber2000first} in which a power spectrum is obtained by correlating between $\mathbf{g}$ and the steering vectors. Accordingly, Rayleigh resolution is defined as the minimum distance between two separable point scatterers using the beamforming method
	\begin{equation}
		\label{Eq_3}
		\rho_{s}=\frac{{\lambda}R_0}{2\Delta b},
	\end{equation} 
	where $\Delta b$ is the total baseline extent. Due to the limitation of baseline extent as a result of the small number of acquisitions, the resolution of the beamforming method is low.
	
	For the detection of the point scatterers, SGLRTC was proposed \cite{pauciullo2012detection}. In this method, detection is performed by finding the maximum of the beamforming power spectrum and a successive cancellation routine is used for finding multiple targets. Despite the high speed of SGLRTC, it has poor resolution.
	
	The ML detector is the optimum solution for the detection of point scatterers with high-resolution capability. In this method, the reflectivity of the $k$ scatterers at $l$-th snapshot is obtained via the least square estimate \cite{kay1993fundamentals}.
	This method which demands high computational load due to the need for a combinatorial search has been used in TomoSAR with different names, nonlinear least square (NLS) \cite{zhu2010very}, sequential generalized likelihood ratio test (SGLRT) \cite{pauciullo2012detection} and support estimation (Sup-GLRT) \cite{budillon2016glrt}. In this article, we chose NLS for naming the ML detector. As a suboptimal method, RELAX algorithm is proposed to reduce the complexity of the NLS estimator \cite{li1996efficient}. In this method, the $k-$ dimensional combinatorial search of NLS is replaced by $k$ one-dimensional search where the covariance matrix is updated in each iteration of the algorithm \cite{gini2002layover}. 
	
	Among different spectral estimation methods, MUSIC has attracted many researchers due to its high resolution capability and moderate complexity. For the MUSIC as a subspace-based method, covariance matrix should be available. Given the equation \eqref{Eq_1} for uncorrelated signal and noise scenario, the covariance matrix of the received data will be as follows
	\begin{equation}
		\label{Eq_4}
		\mathbf{R} = E[\mathbf{g}\mathbf{g}^H] = \sum_{i=1}^{k}\sigma_{s_i}^2 \boldsymbol{a}(s_{m_i})\boldsymbol{a}^H(s_{m_i}) + \sigma_w^2 \mathbf{I}_N,
	\end{equation}  
	where $\sigma_{s_i}^2$ is the statistical mean squared reflectivity of (return power from) the $i$-th target and $\sigma_w^2$ is the noise variance. The exact expression of covariance matrix is not known and must be estimated from the received data. The asymptotic ML estimation of the covariance matrix is obtained from the following equation
	\begin{equation}
		\label{Eq_5}
		\hat{\mathbf{R}} = E[\mathbf{g}\mathbf{g}^H] \cong \frac{1}{L}\sum^L_{l=1}\mathbf{g}(l)\mathbf{g}(l)^H,
	\end{equation}
	which is called SCM. The eigenvalue decomposition of the covariance matrix yields
	\begin{equation}
		\label{Eq_6}
		\mathbf{R} = \begin{bmatrix}
			\mathbf{U_s} & \mathbf{U_n} 
		\end{bmatrix}\begin{bmatrix}
			\mathbf{E_s} & 0 \\
			0 & \mathbf{E_n}
		\end{bmatrix}
		\begin{bmatrix}
			\mathbf{U_s} & \mathbf{U_n} 
		\end{bmatrix}^H,
	\end{equation}
	where $E_s$ and $E_n$ are eigenvalues and $U_s$ and $U_n$ are eigenvectors of signal and noise respectively. The MUSIC power spectrum (pseudo-spectrum) is obtained by the noise subspace as \cite{schmidt1986multiple}
	\begin{equation}
		\label{Eq_7}
		|\pmb{\gamma}|^2 = \frac{1}{\boldsymbol{a}(s_m)^H\mathbf{U}_n\mathbf{U}_n^H\boldsymbol{a}(s_m)}.
	\end{equation}
	
	The MUSIC algorithm requires the model order which is the number of scatterers. In this paper, we assume that the number of scatterers is estimated by one of the methods known from the literature (e.g. the Akaike information criteria (AIC) or minimum description length (MDL) \cite{akaike1974new,rissanen1978modeling} ). 	
	
	Due to the influence of the covariance matrix estimation on the performance of the MUSIC algorithm, it is desired to apply a consistent estimator for the covariance matrix. The accuracy of SCM is degraded in sample starved conditions where the number of snapshots is comparable with the array dimension. The estimation accuracy of the covariance matrix increases by an increasing number of snapshots, however, the number of looks in SAR tomography is limited. Additionally, an increasing number of looks deteriorate the quality of SAR
	image in range-azimuth directions. Thus, high-quality methods are required for the covariance matrix estimation. Among the various solutions for covariance matrix estimation, the correlation subspace method is considered due to its appealing properties. 		\section{Covariance matrix estimation: the correlation subspace method}\label{sec3}		
	This method starts with finding a specific subspace which is used as a constraint for estimation of the covariance matrix. It is shown that vectorizing the covariance matrix (after removal of the noise variance), results in a vector that lies in a subspace called correlation subspace \cite{rahmani2016subspace}, i.e,
	\begin{equation}
		\label{Eq_8}
		\mathrm{vec} \left(\mathbf{R} - \sigma_n^2 \mathbf{I}_N \right) \in \mathcal{CS}.
	\end{equation} 
	Irrespective of definition of the correlation subspace, it is obtainable by the expression below
	\begin{equation}
		\label{Eq_9}
		\mathcal{CS}=\mathrm{col}(\mathbf{Q}),
	\end{equation} 
	where $\mathrm{col}(\mathbf{Q})$ is the column space of the $\mathbf{Q}$ matrix. Further details about definition of correlation subspace and the derivation of $\mathbf{Q}$ matrix is noted in \ref{App1}. Equations \eqref{Eq_8} and \eqref{Eq_9} state that $\mathrm{vec} \left(\mathbf{R} - \sigma_n^2 \mathbf{I}_N \right)$ is spanned by the column vectors of $\mathbf{Q}$ matrix, or equivalently it is in the form of $\sum_{p=1}^Pc_p\mathbf{Q}(:,p)$ where $c_p$ is a complex number and $P$ is dimension of $\mathbf{Q}$.
	
\begin{figure}
		\centering
		\includegraphics[width=0.75\linewidth]{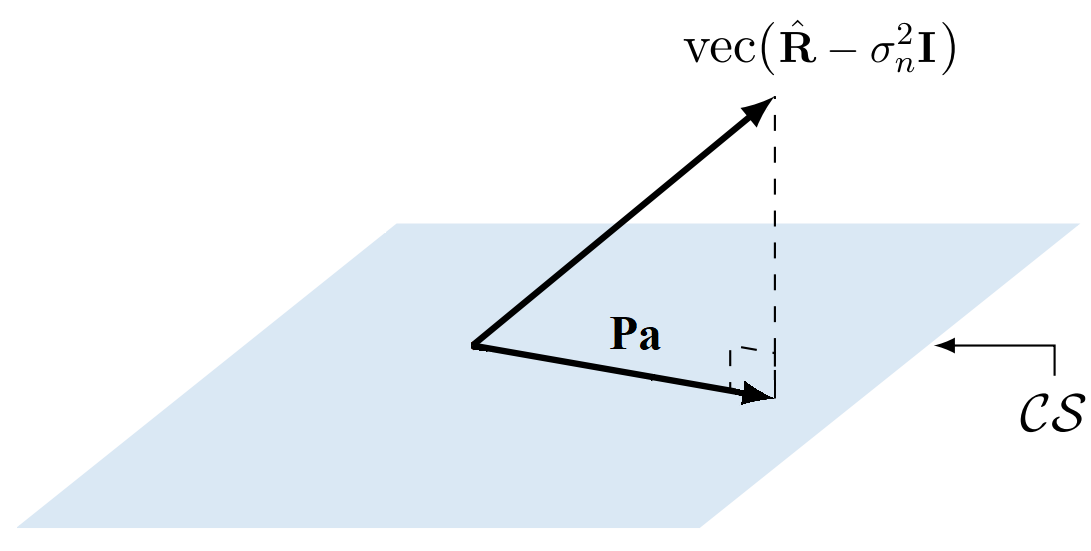}
		\caption{Orthogonal projection on the correlation subspace as the solution of the optimization problem Pa \cite {liu2017correlation}.}
		\label{fig:orthogonal-projection}
	\end{figure}

	In the next step,  $\mathbf{Q}$ is employed as a constraint for covariance matrix estimation. For this, an optimization problem is considered that fits the covariance matrix to the de-noised sample covariance matrix, i.e. finds the closest vector to $\mathrm{vec} \left(\hat{\mathbf{R}} - \sigma_n^2 \mathbf{I}_N \right)$. On the other hand, this vector lies in the $\mathcal{CS}$ which is the subspace spanned by the $\mathbf{Q}$ matrix. As shown in Fig. \ref{fig:orthogonal-projection}, the closest vector to the $\mathrm{vec} \left(\hat{\mathbf{R}} - \sigma_n^2 \mathbf{I}_N \right)$ which also lies in the correlation subspace is its orthogonal projection to the correlation subspace. Equivalently, the orthogonal complement projection of it to the correlation subspace is zero. Hence, the covariance matrix is estimated using the following optimization problem $\mathrm{(P)}$ \cite{rahmani2016subspace}.
	\begin{equation}
		\label{Eq_10}
		\mathrm{(P)} \mathrm{:} \enspace \mathbf{R}_{\mathrm{P}}=\arg\min_{\mathbf{R}} \|\hat{\mathbf{R}} - \sigma_n^2\mathbf{I} - \mathbf{R}\|_2^2
	\end{equation} 
	\begin{equation}
		\label{Eq_11}
		\quad \quad \quad \quad \quad \quad \quad \quad \quad \mathrm{s.t.} \quad (\mathbf{I}-\mathbf{Q}\mathbf{Q}^{\dagger})\mathrm{vec} \left(\mathbf{R}\right) = \mathbf{0},
	\end{equation}
	\begin{equation}
		\label{Eq_12}
		\quad \quad \quad \quad \quad \mathbf{R} \succcurlyeq \mathbf{0},
	\end{equation}	
	where $\mathbf{Q}^{\dagger}$ is the pseudo-inverse of the $\mathbf{Q}$ matrix. The second constraint \eqref{Eq_12} states the semi-positive definiteness of the covariance matrix. The sample covarianve matrix is always positive semidefinite, but  enforcing the sonstraint Due to the high cost of the numerical optimization for this solution, a suboptimal method was presented where the optimization problem $\mathrm{(P)}$ is approximated by the following two sub-problems \cite{rahmani2016subspace}	
	\begin{equation}
		\label{Eq_13}
		\mathrm{(Pa)} \mathrm{:} \enspace \mathbf{R}_{\mathrm{Pa}}=\arg\min_{\mathbf{R}} \|\hat{\mathbf{R}} - \sigma_n^2\mathbf{I} - \mathbf{R}\|_2^2
	\end{equation} 
	\begin{equation}
		\label{Eq_14}
		\hspace{0.1cm} \quad \quad \quad \quad \quad \quad \quad \quad \quad \mathrm{s.t.} \quad (\mathbf{I}-\mathbf{Q}\mathbf{Q}^{\dagger})\mathrm{vec} \left(\mathbf{R}\right) = \mathbf{0},
	\end{equation}
	
	\begin{equation}
		\label{Eq_15}
		\mathrm{(Pb)}  \mathrm{:} \enspace \mathbf{R}_{\mathrm{Pb}}=\arg\min_{\mathbf{R}} \|\mathbf{R}_{\mathrm{p_a}} - \mathbf{R}\|_2^2 \hspace{0.7cm}
	\end{equation}
	\begin{equation}
		\label{Eq_16}
		\hspace{0.1cm} \quad \quad \mathrm{s.t.} \quad \mathbf{R} \succcurlyeq \mathbf{0}.
	\end{equation}
	
	The solution of sub-problem $\mathrm{(Pa)}$ is the orthogonal projection of  $\mathrm{vec} \left(\hat{\mathbf{R}} - \sigma_n^2 \mathbf{I}_N \right)$ onto the subspace spanned by $\mathbf{Q}$ which is analytically derived as \cite{meyer2000matrix,abdi2007method}
	\begin{equation}
		\label{Eq_17}
		\mathrm{vec} \left(\mathbf{R}_{\mathrm{Pa}}\right)=\mathbf{Q}\mathbf{Q}^{\dagger}\mathrm{vec} \left(\hat{\mathbf{R}} - \sigma_n^2 \mathbf{I}_N \right).
	\end{equation}	
	The second sub-problem could be solved easily by only considering the positive eigenvalues of $\mathbf{R}_{\mathrm{Pa}}$ and their corresponding eigenvectors
	\begin{align}
		\label{Eq_18}
		\mathbf{R}_{\mathrm{P}}	&= \mathbf{R}_{\mathrm{Pb}}=\sum_{i=1}^{q}\lambda_i \mathbf{v}_i \mathbf{v}_i^H,
	\end{align}	
	where $\lambda_i$ are non-negative eigenvalues and $\mathbf{v}_i$ are their corresponding eigenvectors of the eigendecomposition of $\mathbf{R}_{\mathrm{Pa}}$. The procedure of suboptimal covariance matrix estimation based on correlation subspace is given in Algorithm \ref{alg.CORRSUB}.
	
	The early simulations show significant superiority of the correlation subspace method over the classical MUSIC and also the SPICE \cite{stoica2010spice}, which is a well-known method in the MMV literature. In addition, the computational complexity of the suboptimal correlation subspace is much lower than the SPICE method which needs a large number of matrix inversion to achieve an acceptable result.

	\begin{algorithm}[t!] 
		\caption{Covariance matrix estimation using correlation subspace} 
		\label{alg.CORRSUB} 
		\input{algs/CORRSUB}
	\end{algorithm}

	\section{Sequential MUSIC algorithms}\label{sec4}
	In the classical MUSIC method, the result of beamforming is a power spectrum in which, the local maxima (peak points) indicate the position of the targets. Although the peak points are visually recognizable in the power spectrum, their selection requires a peak detecting or peak-picking approach. Sequential or recursive MUSIC methods do this in a successive cancellation routine. The main procedure in all of these methods is finding the strongest component of the power spectrum. Then, after removing the related component of the first target through orthogonal complement projection, the second target is detected in the updated power spectrum. Although there are slight differences between the different versions of the sequential MUSIC algorithms, their routine is similar. In this article, we choose RAP-MUSIC among these methods due to its wider applications in source localization \cite{jatoi2014survey,golub2003separable}. 
	
	The method begins with locating the first target by finding the maximum of power spectrum based on the signal subspace as follows
	\begin{equation}
		\label{Eq_19}
		m_1 = \arg\max_{i}\{\boldsymbol{a}(s_i)^H\mathbf{U}_s\mathbf{U}_s^H\boldsymbol{a}(s_i)\},
	\end{equation} 	
	where $m_1$ is the index of the first target. In the next step, the orthogonal complement projection is applied as a cancelling factor on the main power spectrum to find the location of second target
	\begin{equation}
		\label{Eq_20}
		m_2 = \arg\max_{i}\{\boldsymbol{a}(s_m)^H(\mathbf{I}_N - \Pi_{m_1})\mathbf{U}_s\mathbf{U}_s^H(\mathbf{I}_N - \Pi_{m_1})\boldsymbol{a}(s_m)\},
	\end{equation}
	where $\Pi_{m_1}$ is the orthogonal projection to the subspace extended by $\boldsymbol{a}(s_{m_1})$ and is defined as
	\begin{equation}
		\label{Eq_21}
		\Pi_{m_1} = \boldsymbol{a}(s_{m_1})\boldsymbol{a}[(s_{m_1})^H\boldsymbol{a}(s_{m_1})]^{-1}\boldsymbol{a}(s_{m_1})^H.
	\end{equation}
	\begin{algorithm}[t!] 
		\caption{RAP-MUSIC method} 
		\label{alg.RAPMUSIC} 
		\input{algs/RAPMUSIC}

	\end{algorithm}	
	The routine of RAP-MUSIC is shown in Algorithm \ref{alg.RAPMUSIC}. Despite the advantages mentioned, the accuracy of this method is not high enough. Hence, it is desirable to design a sequential MUSIC algorithm with a higher resolution power. That's why the RCC-MUSIC method is suggested, which is introduced in the following.	
	\subsection{RCC-MUSIC: the proposed method}
	Considering equation \eqref{Eq_4}, it is seen that the covariance matrix is a linear combination of some matrices in the form $\boldsymbol{a}(s_{m_i})\boldsymbol{a}^H(s_{m_i})$ whose coefficients are the square of reflectivities of the targets. Accordingly, a cancellation routine is to estimate the reflectivity of a detected target and subtract its related component from the covariance matrix, i.e.
	\begin{equation}
		\label{Eq_22}
		\mathbf{R}^{(1)} = \mathbf{R} - \hat{\sigma}_{s_1}^2\boldsymbol{a}(s_{m_1})\boldsymbol{a}^H(s_{m_1}),
	\end{equation}	
	where $m_1$ is the location of the first target and $\hat{\sigma}_{s_1}^2$ is its estimated squared reflectivity. The squared reflectivity is estimated by averaging over the estimated squared reflectivities of the individual snapshots via NLS method, i.e.
	\begin{align}
		\label{Eq_23}
		\hat{\sigma}_{s_1}^2 &=E\{|\hat{\pmb{\gamma}}_{\Omega^{1}}|^2\} \notag\\	
		&= \frac{1}{L}\sum_{l=1}^{L}|(\mathbf{A}^H_{\Omega^{1}}\mathbf{A}_{\Omega^{1}})^{-1}\mathbf{A}^H_{\Omega^{1}}\mathbf{g}(l)|^2\notag\\
		&= \frac{1}{L}\sum_{l=1}^{L}|(\boldsymbol{a}^H(s_{m_1})\boldsymbol{a}(s_{m_1}))^{-1}\boldsymbol{a}^H(s_{m_1})\mathbf{g}(l)|^2,
	\end{align}
	where $\Omega^{1}$ is the location of targets in the first iteration ($\Omega^{1}=m_1$), and consequently, $\mathbf{A}_{\Omega^{1}}=\boldsymbol{a}(s_{m_1})$. Next, using the signal subspace derived from the updated covariance matrix $\mathbf{R}^{(1)}$, the new power spectrum is obtained for detection of the next target. The procedure of the RCC-MUSIC method is shown in Algorithm \ref{alg.RCCMUSIC}. In this algorithm, $\Omega^{i}$ is the location of the targets in $i$-th iteration and $\Lambda$ is the vector of the estimated squared reflectivities of the detected targets. $\mathbf{R^{(i)}}$ is the covariance matrix in $i$-th iteration, from which the related components of the previously detected targets have been removed. 
	
	Similar to RAP-MUSIC, the total number of cancellations in RCC-MUSIC is equal to the number of targets minus one. The only difference between RCC-MUSIC and the previous sequential MUSIC algorithms is that in each iteration of RCC-MUSIC, a further eigendecomposition is required to obtain the updated noise and signal subspaces.
	
	\begin{algorithm}[t!] 
		\caption{RCC-MUSIC method} 
		\label{alg.RCCMUSIC} 
		\input{algs/RCCMUSIC}
	\end{algorithm}		
	
	In section \ref{sec6} the impact of covariance matrix estimation on the performance of the MUSIC method will be demonstrated. Due to the advantage of the correlation subspace method, we combine it with the proposed RCC-MUSIC and investigate the resulting improvement through simulation results.
	\begin{figure}
		\centering
		\includegraphics[width=0.87\linewidth]{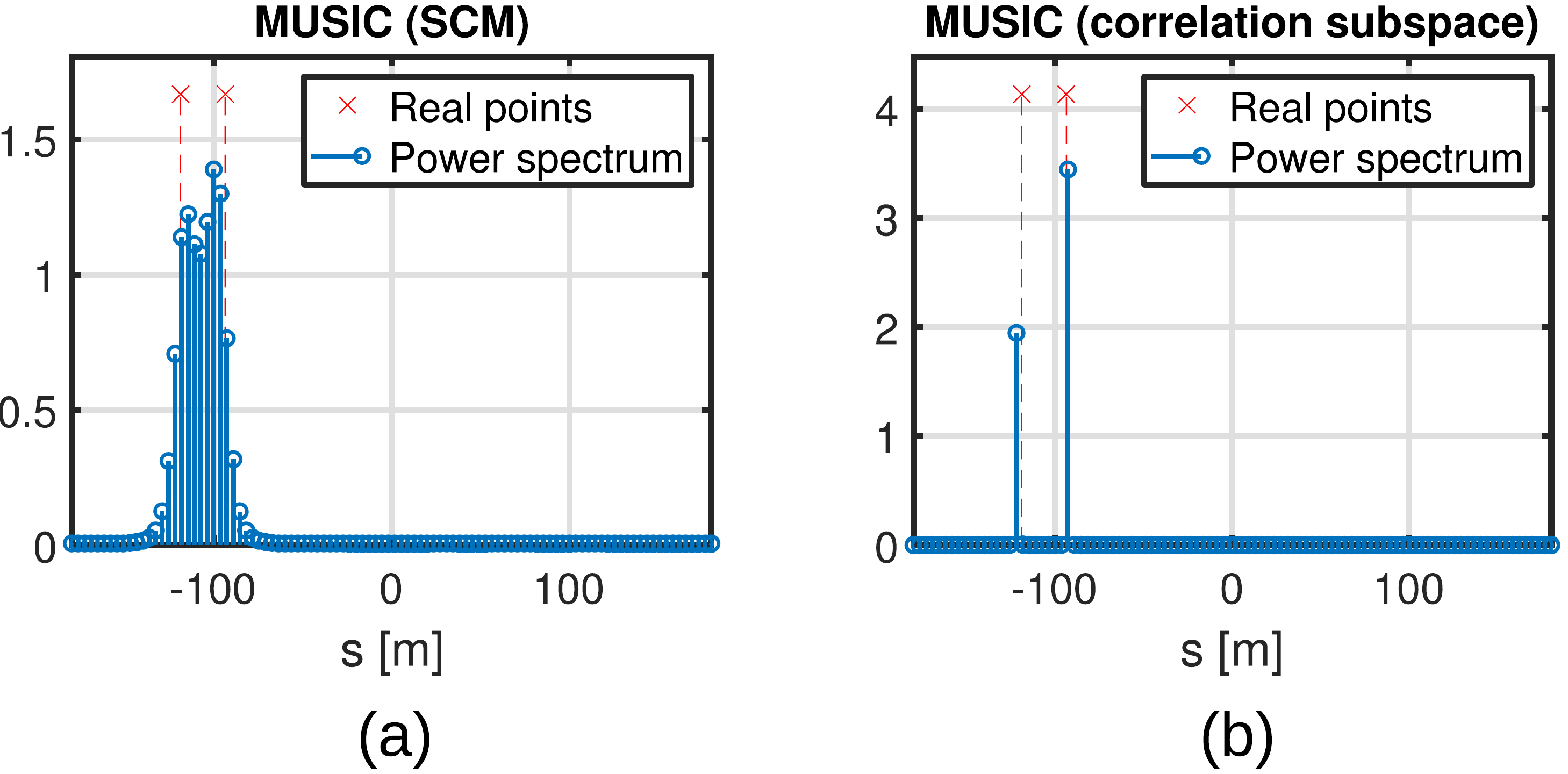}
		\caption{Comparison of covariance matrix estimation methods in classical MUSIC algorithm (a) sample covariance matrix (SCM) (b) correlation subspace for SNR = 0 dB and $\alpha=0.7$.}
		\label{fig:covscmcorrsubcompare-1}
	\end{figure}
	\section{Performance evaluation}\label{sec5}
	In this section, the efficiency of the RCC-MUSIC method is investigated and its performance will be compared with the methods including NLS, SGLRTC, RELAX, and RAP-MUSIC. The performance of these methods is evaluated by two criteria of estimation accuracy and computational complexity. 
	
	In the numerical simulation, two scatterers are considered ($k=2$) in each range-azimuth cell, and the number of snapshots is set to $L = 25$. SNR per scatterer is defined as the mean squared reflectivity of a scatterer to the noise power. In the double scatterers scenario, amplitudes of the both scatterers are considered as $\sigma_s$ and hence, SNR = $\frac{\sigma^2_s}{\sigma^2_w}$. In addition, the distance of two targets which is normalized to Rayleigh resolution is shown by $\alpha$ and is defined as $\alpha=\frac{1}{\rho_s}(s_{m_2}-s_{m_1})$.
	Other parameters chosen for the simulation are the number of antennas $N=14$, Rayleigh resolution $\rho_s=26$ m and the number of grids in elevation $M=234$. To measure the estimation error, root mean square error (RMSE) of the estimated location of point targets is considered
	\begin{equation}
		\label{Eq_24}
		RMSE=\sqrt{\frac{1}{N_{Det}}\sum\limits_{i=1}^{N_{Det}}{\frac{(s_{m_1}-\hat{s}_{m_{1i}})^2+(s_{m_2}-\hat{s}_{m_{2i}})^2}{2}}},
	\end{equation}
	where $N_{Det}$ is the total number of the detections, $s_{m_1},s_{m_2}$ are real positions of targets in elevation and $\hat{s}_{m_{1i}},\hat{s}_{m_{2i}}$ are estimated positions of two targets at $i$-th successful detection. 
	
	Cramer-Rao lower bound (CRLB) is the asymptotic bound for for variance of an unbiased estimator. In SAR tomography, the CRLB expression for estimation of a scatterer's location in the single scatterer case is as follows \cite{rife1974single,lee1992cramer}
	\begin{equation}
		\label{Eq_25}
		\hspace{1.1cm} CRLB_{1} = \frac{3}{2\pi^2}(\frac{\rho_s}{\sqrt{L.N.SNR}})^2, \hspace{0.83cm} k=1,\\
	\end{equation}		
	where $CRLB_{1}$ stands for the cramer-rao bound in the single scatterer case. For double scatterers, deriving the CRLB is more complex comparing to that of the single scatterer. This is due to the lengthy expressions appear in calculation of inverse of Fisher information matrix (FIM) which is a function of distance of two targets and their phase difference. A straitforward approximation for CRLB in double-scatterer case is derived as \cite{lee1992cramer, swingler1993frequency}
	\begin{equation}
		\hspace{1.1cm} CRLB_{2} \approx CRLB_{1}.\upzeta(\alpha,\Delta{\phi}), \hspace{0.95cm} k=2
		\label{Eq_26}
	\end{equation}  
	where $\upzeta(\alpha)$ is the normalized CRLB defined as
	\begin{equation}
		\upzeta(\alpha)\approx \max\{\frac{15}{\pi^2}\alpha^{-2},1\}
		\label{Eq_27}
	\end{equation}
	
	Equations \eqref{Eq_26} and \eqref{Eq_27} show that for two closely-spaced scatterers, the CRLB increases by a factor which is a function of two targets' distance. The closed-form relationship \eqref{Eq_26}-\eqref{Eq_27} is verified numerically which asserts its validity for low and moderate SNR values. For a fair comparison between RMSE and CRLB of one scatterer, the MSE of the two scatterers is averaged by a division by 2 in \eqref{Eq_24}.
	
	Before quantitative comparison of the results, it is worth evaluating them qualitatively through sketching their power spectrum.
	\begin{figure}
		\centering
		\includegraphics[width=1.15\linewidth]{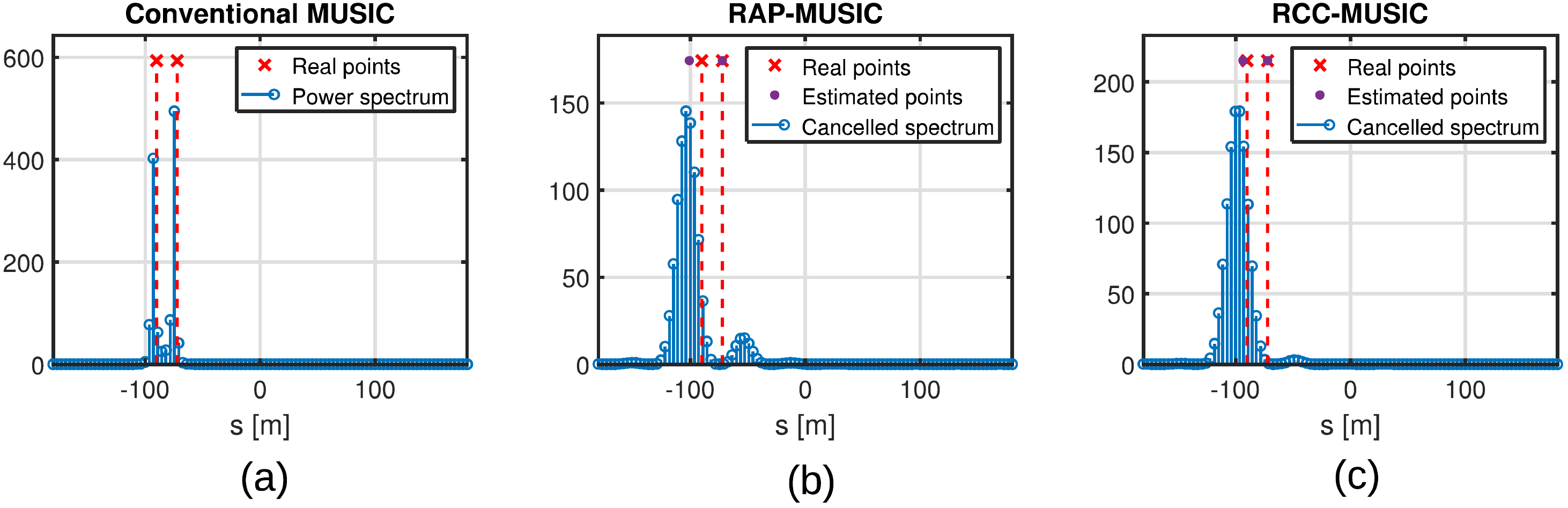}
		\caption{Comparison of MUSIC methods (a) classical MUSIC (b) RAP-MUSIC (c) RCC-MUSIC for SNR = 14 dB and $\alpha=0.5$.}
		\label{fig:sequentialmusicpowspectsnr14spc0}
	\end{figure}
	In Fig. \ref{fig:covscmcorrsubcompare-1}, the result of estimating the covariance matrix using the correlation subspace is shown for the MUSIC algorithm at SNR = 0 dB and $\alpha=0.7$. It is observed that employing the improved covariance matrix estimator, increases the resolution power as the two merged targets become separable.
	
	In Fig. \ref{fig:sequentialmusicpowspectsnr14spc0}, the performance of RAP-MUSIC and RCC-MUSIC methods are compared in estimating the location of the targets at SNR = 14 dB and $\alpha=0.5$. As shown in [Fig. \ref{fig:sequentialmusicpowspectsnr14spc0}(a)], the two peaks of the MUSIC power spectrum are visually distinguishable and almost correspond to the actual location of the targets. Also, in [Fig. \ref{fig:sequentialmusicpowspectsnr14spc0}(b)] and [Fig. \ref{fig:sequentialmusicpowspectsnr14spc0}(c)], the power spectrum of RAP-MUSIC and RCC-MUSIC are displayed after canceling the related component of the first target. It can be seen that the estimated location of the second target for the RCC-MUSIC method is closer to the real locations compared with RAP-MUSIC.
	
\begin{figure}
		\centering
		\includegraphics[width=1.15\linewidth]{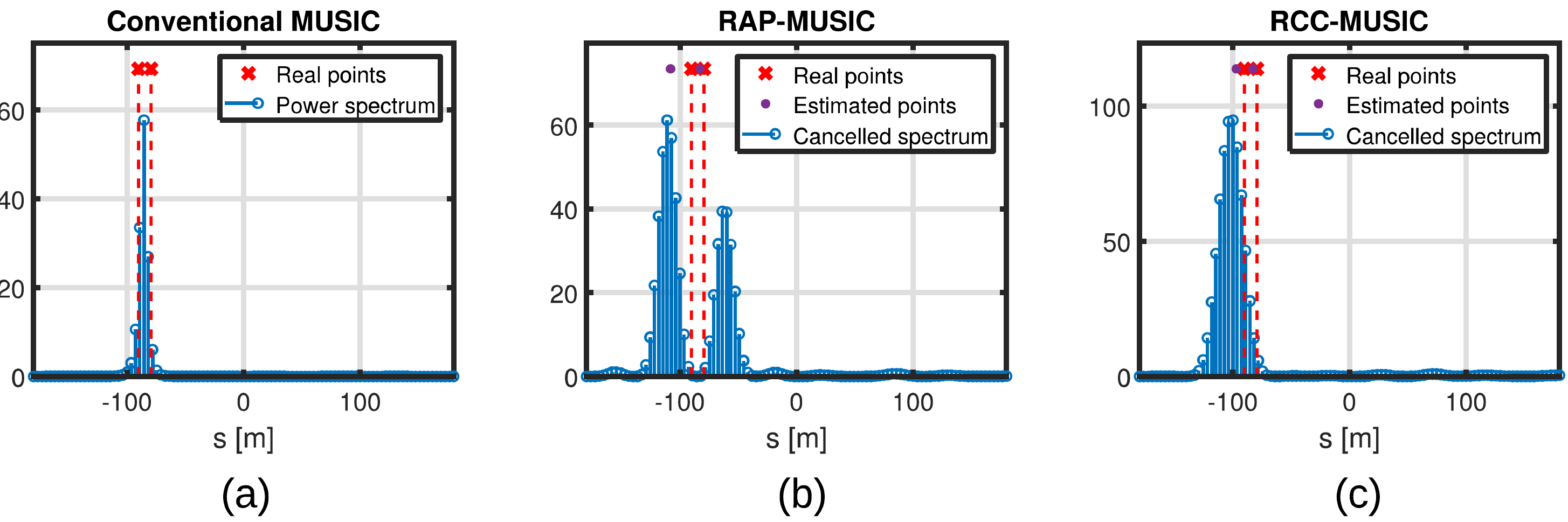}
		\caption{Comparison of MUSIC methods (a) classical MUSIC (b) RAP-MUSIC method (c) RCC-MUSIC method for SNR = 8 dB and $\alpha=0.3$.}
		\label{fig:sequentialmusicpowspectsnr8spc0}
	\end{figure}

	Fig. \ref{fig:sequentialmusicpowspectsnr8spc0} shows the performance of the methods for low SNR conditions and closely-spaced targets where SNR = 8 dB and $\alpha=0.3$. Noteworthy, both RMSE and CRLB values are scaled to Rayleigh resolution $\rho_s$. As can be seen, the classical MUSIC method could not resolve the two targets, while by sequential MUSIC methods, two distinct points are given. However, due to the inaccuracy of the location of the first target, the position of the second target is also estimated by error. The results show that the RCC-MUSIC method still has a significant advantage over the RAP-MUSIC method.
	
\begin{figure}
		\centering
		\includegraphics[width=1\linewidth]{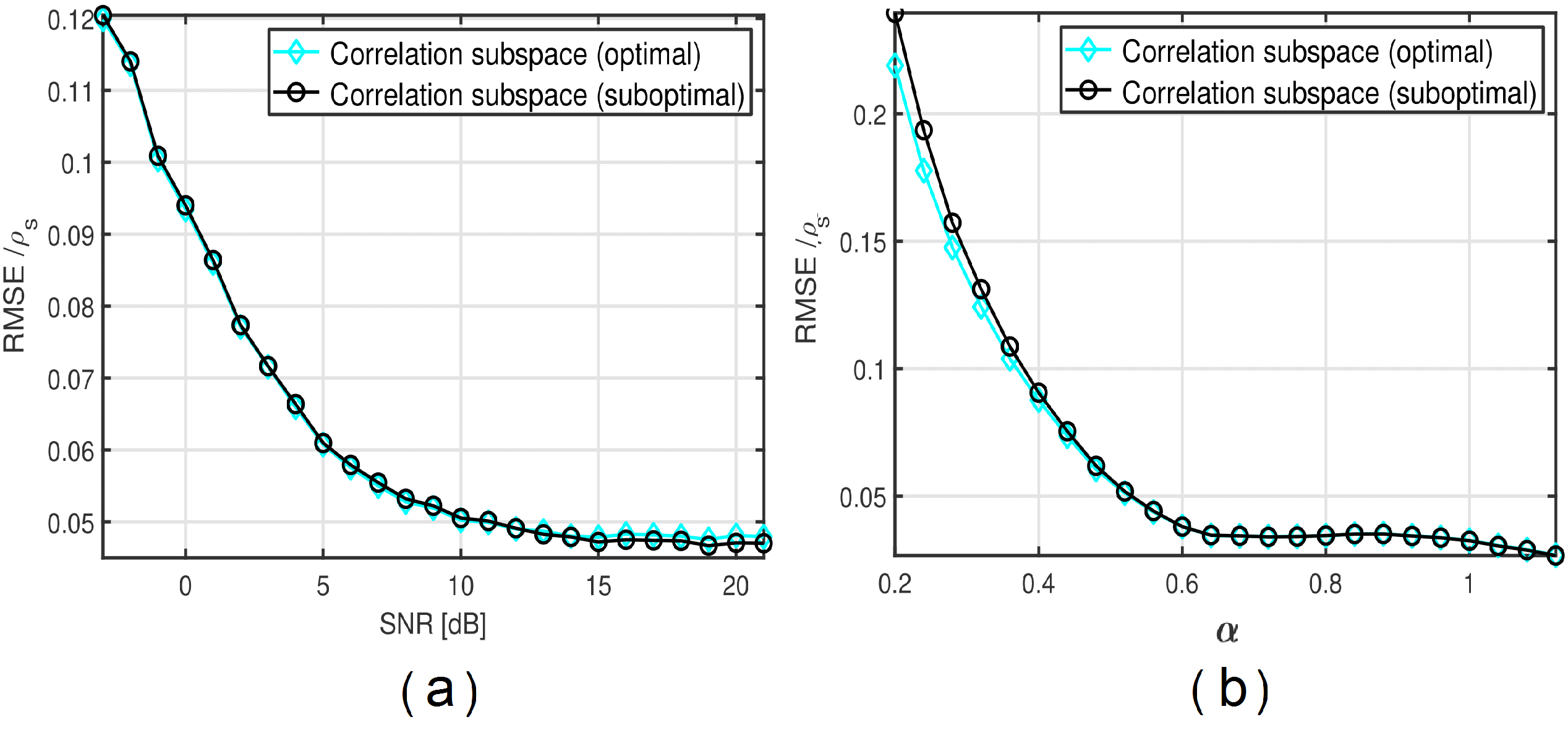}
		\caption{Comparison of optimal and suboptimal solutions in implementing the correlation subspace method (a) versus SNR (b) versus distance of targets.}
		\label{fig:corrsubvssnr-1}
	\end{figure}

	Fig. \ref{fig:corrsubvssnr-1} measures the estimation accuracy of RCC-MUSIC using optimal and suboptimal solutions of the correlation subspace versus of SNR and $\alpha$. As seen in Fig. \ref{fig:corrsubvssnr-1}(a), the accuracy of the suboptimal solution is almost equal to that of the optimal solution. The results of Fig. \ref{fig:corrsubvssnr-1}(b) show that the difference between the optimal and suboptimal solutions is negligible and can be tolerated. Also, the two solutions, have been compared in terms of computational complexity. In both solutions, the Q matrix is calculated offline at the beginning of processing. For the suboptimal solution, just a simple matrix multiplication and an eigendecomposition are needed, while the optimal solution requires convex optimization solvers. Based on the experiments, the suboptimal method is significantly faster than the optimal method. As a result, the suboptimal method, while maintaining the acceptable quality is the best option for implementing the correlation subspace method.
	
\begin{figure}
		\centering
		\includegraphics[width=0.81\linewidth]{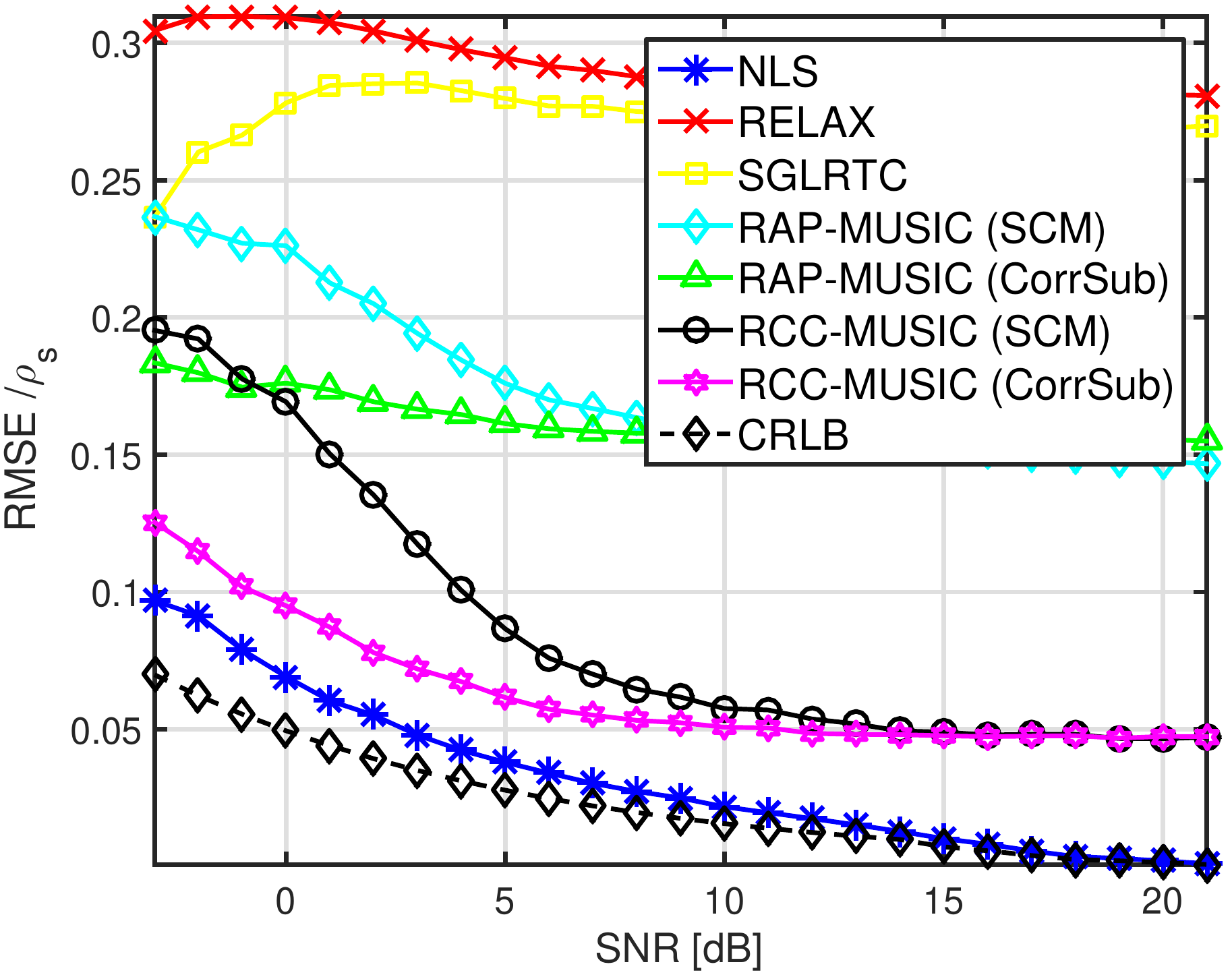}
		\caption[]{Estimation error as a function of SNR, for $\alpha=0.5$.}
		\label{fig:msevssnrmmv}
	\end{figure}

	In the next experiment, in Fig. \ref{fig:msevssnrmmv}, the estimation accuracy of the methods has been measured in terms of SNR. In this figure, the covariance matrix estimation methods based on the sample covariance matrix and correlation subspace have been shown with SCM and CorrSub, respectively. The highest estimation error belongs to the SGLRTC method while the NLS method has the least error. Comparing the estimation methods, it can be seen that the error of the RCC-MUSIC method is about $0.1\rho_s$ and $0.23\rho_s$ lower than RAP-MUSIC and SGLRTC, respectively. Also, it is observed that applying the correlation subspace in covariance estimation, brings about up to 6 dB SNR gain.
	
	Considering the results of Fig. \ref{fig:msevssnrmmv}, it is observed that the error of SGLRTC and sequential MUSIC algorithms do not converge to zero even when SNR tends to the infinity and there is an error floor. What is common in all of these methods, is performing the component cancellation of the data, unlike the NLS, wherein the positions of all targets are estimated simultaneously. A justification is that by removing a part of the data, although the position of the next component is obtained, some information is also lost. The point here is that under normal circumstances, the SNR value is low or medium, and using the successive cancellation routine is a favorite solution. The RCC-MUSIC method based on correlation subspace covariance matrix estimation is superior to other sequential methods.
	
\begin{figure}
		\centering
		\includegraphics[width=0.82\linewidth]{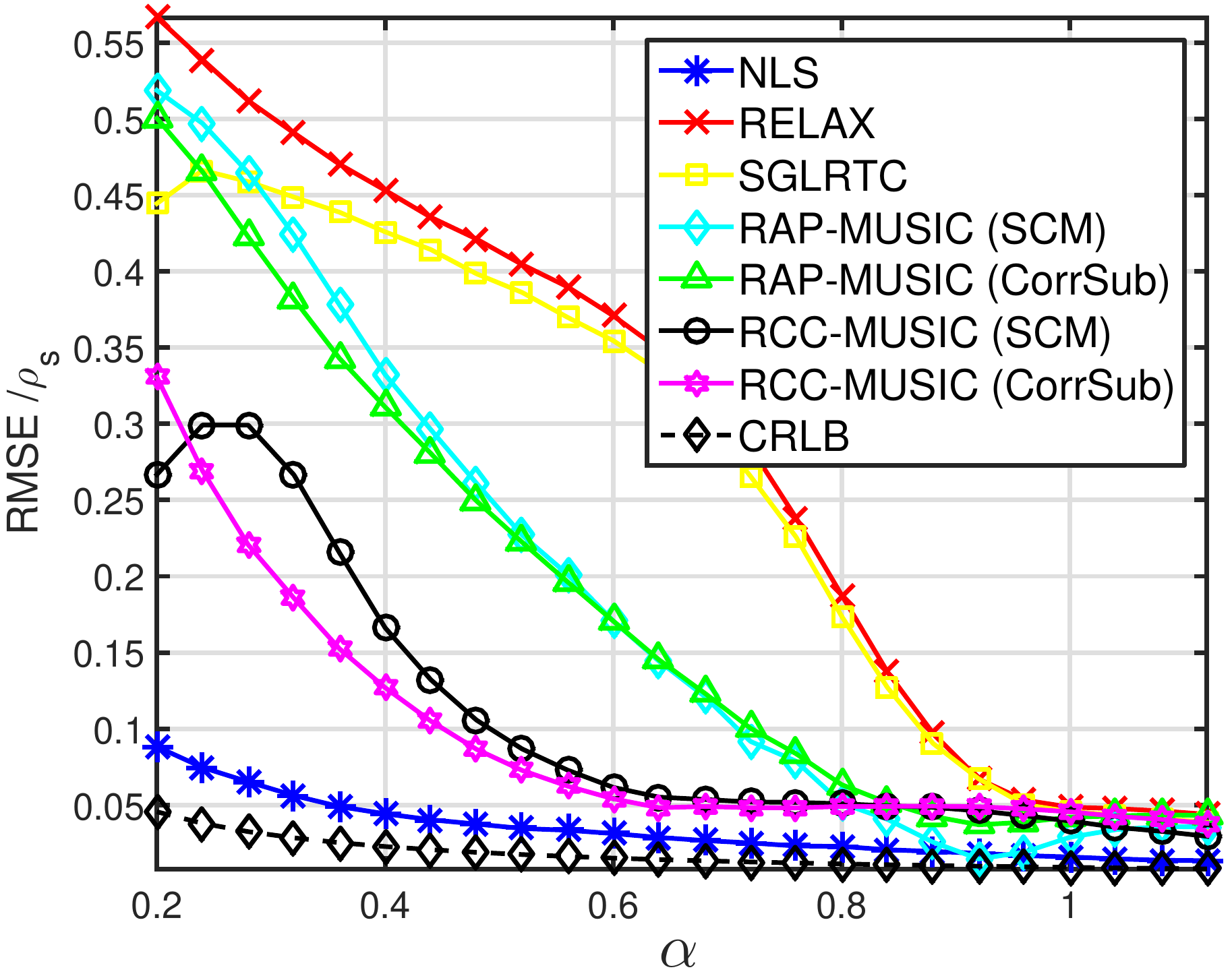}
		\caption[]{Estimation error as a function of distance of the targets, SNR = 9 dB.}
		\label{fig:msevsspcmmv}
	\end{figure}

	In Fig. \ref{fig:msevsspcmmv}, the RMSE of different methods is sketched versus the distance of targets. The results show that the RCC-MUSIC method decreases the RMSE up to about $0.15\rho_s$ relative to the RAP-MUSIC. Also, at the SNR = 9 dB, the correlation subspace makes a slight improvement, maximally up to $0.05\rho_s$ decrease in RMSE.	
	\begin{table}[ht]
		\caption{Computational complexity for different methods.} 
		\label{tab:Elapsed}
		\begin{center}       
			\begin{tabular}{|l||l|} 
				\hline
				\rule[-1ex]{0pt}{3.5ex}  Method & Complexity  \\
				\hline\hline\rule[-1ex]{0pt}{3.5ex}  SGLRTC & $\mathcal{O}((2N^2+2N)M)+N^3+(L+1)N^2$ \\
				\hline
				\rule[-1ex]{0pt}{3.5ex} RELAX & $\mathcal{O}((2N^2+2N)M)+2LN^2$\\
				\hline
				\rule[-1ex]{0pt}{3.5ex}  RAP-MUSIC (SCM) & $\mathcal{O}((N^3+4N^2+2N)M+N^3+(L+1)N^2)$ \\
				\hline
				\rule[-1ex]{0pt}{3.5ex}  RAP-MUSIC (CorrSub) & $\mathcal{O}((N^3+4N^2+2N)M+N^4+N^3+(L+1)N^2)$ \\
				\hline
				\rule[-1ex]{0pt}{3.5ex}  RCC-MUSIC (SCM) & $\mathcal{O}((2N^3-2N^2+2N)M+2N^3+LN^2+(2L+1)N)$\\
				\hline
				\rule[-1ex]{0pt}{3.5ex}  RCC-MUSIC (CorrSub) & $\mathcal{O}((2N^3-2N^2+2N)M+N^4+2N^3+LN^2+(2L+1)N)$\\
				\hline
				\rule[-1ex]{0pt}{3.5ex}  NLS & $\mathcal{O}((\frac{N^3}{2}+N^2+4N)(M^2-M)+LN^2)$\\
				\hline
				
			\end{tabular}
		\end{center}
	\end{table} 
	
	\subsection{Computationl complexity}\label{sec5}
\begin{figure}
		\centering
		\includegraphics[width=0.79\linewidth]{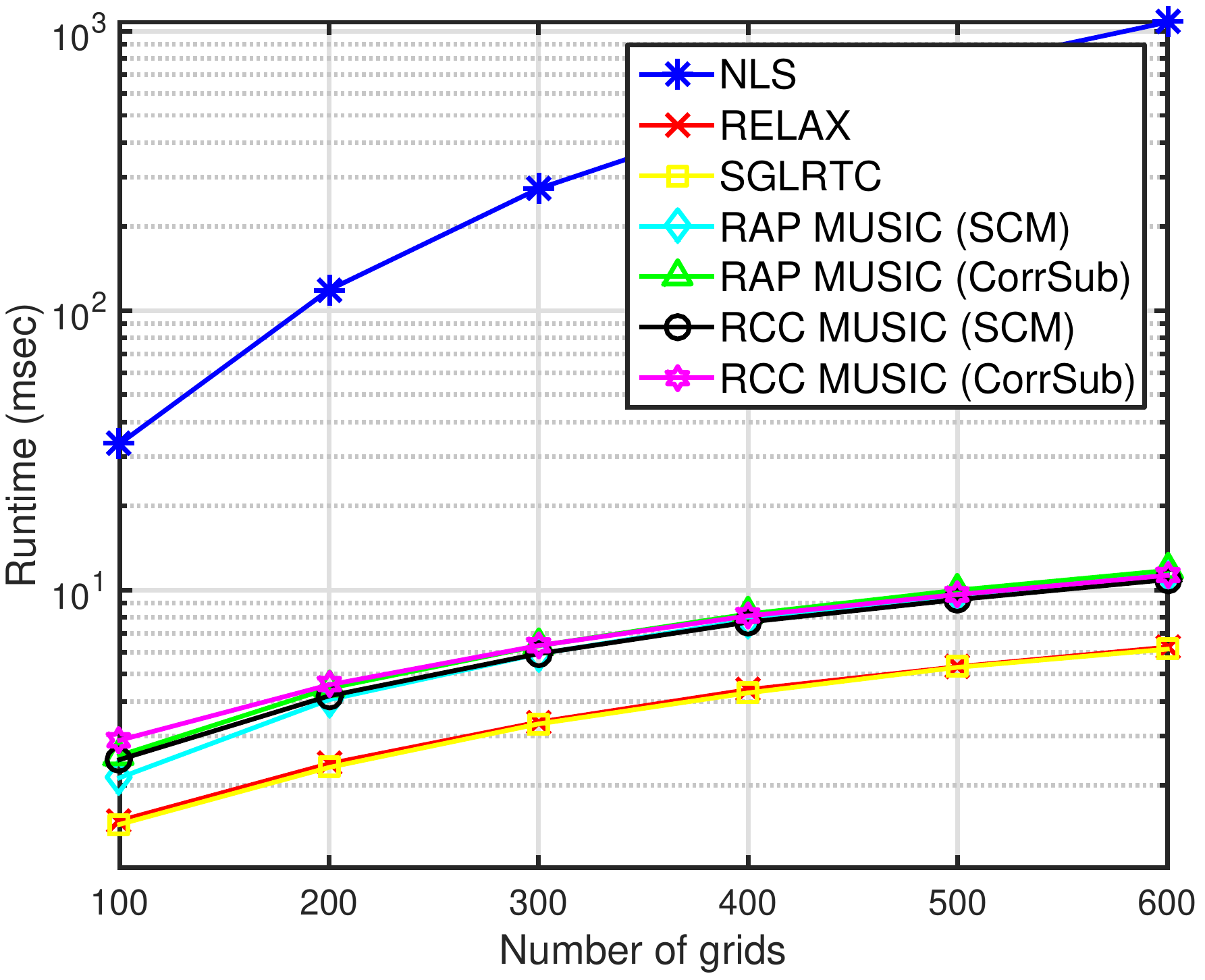}
		\caption{Comparison of elapsed time for different methods versus number of grids (milliseconds).}
		\label{fig:timerccmusiccomparemmv}
	\end{figure}

		After assessment of the accuracy, the complexity of the methods is to be investigated. For all of the mentioned methods, the calculation of the sample covariance matrix is the first step. For the MUSIC-based methods, eigendecomposition is the next processing. Also, all of the methods deal with the power spectrum peak-finding, i.e. MUSIC-based, SGLRTC and RELAX which require a spectrum grid search. The result of the complexity analysis is shown in table \ref{tab:Elapsed} which is obtained regarding the number of complex multiplications required for the methods. For numerical verification of the complexity, the runtime of different methods is sketched versus $M$ in Fig. \ref{fig:timerccmusiccomparemmv}. 
		It is worth saying that the NLS method is implemented using a complexity-reduced version of NLS called CA-NLS \cite{naghavi2020super}. However, as shown in Fig. \ref{fig:timerccmusiccomparemmv}, NLS is the most complex method. The reason is according to table \ref{tab:Elapsed} where the complexity of the NLS is proportional to $\mathcal{O}(M^2)$, but it is proportional to $\mathcal{O}(M)$ for the other methods. The MUSIC-based methods are moderately complex and SGLRTC and RELAX have the least complexity. Besides, it is shown that for the considered $N = 14$, using the correlation subspace has a negligible effect on the complexity. 
		
\begin{figure}
		\centering
		\includegraphics[width=0.79\linewidth]{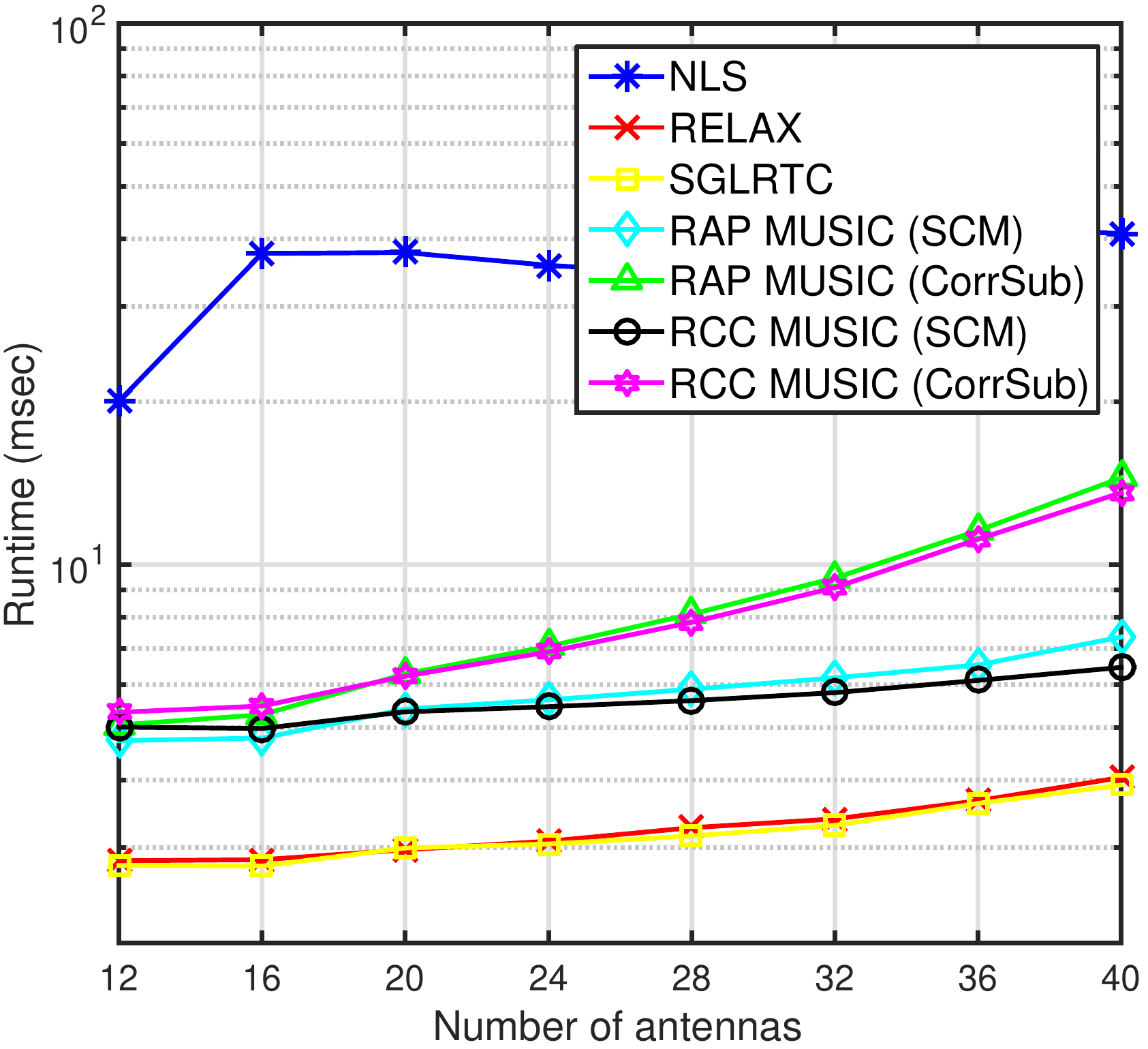}
		\caption{Comparision of elapsed times of the methods versus number of antennas}
		\label{fig:timecomparevsn}
	\end{figure}

		For a more detailed comparison, the complexity of the methods is investigated regarding the number of antennas ($N$) as plotted in Fig. \ref{fig:timecomparevsn}. It is seen that increasing $N$ makes a faster increase in runtime for the MUSIC methods using correlation subspace rather than SCM. This is due to the matrix multiplication in equation \ref{Eq_17} which leads to an extra $N^4$ term for correlation subspace methods as noted in table \ref{tab:Elapsed}. An interesting point is that the complexity of RAP-MUSIC is a bit more than RCC-MUSIC. It shows that the complexity of extra eigendecomposition utilized in the RCC-MUSIC method is generally lower than that of multiplications required for projection in RAP-MUSIC. A justification is that the eigendecomposition of the $N \times N$ covariance matrix which is supposed to be $\mathcal{O}(N^3)$, can be implemented with lower complexity due to the hermitian structure of the covariance matrix and using fast algorithms. 
		
		Therefore, if a very high resolution is desired, the NLS method is suitable. Otherwise, for regular conditions where the number of antennas is not very high, the RCC-MUSIC method using the correlation subspace is the preferred option.	
	
	\subsection{Notes on covariance matrix subspace estimation}
	By examining the equation \eqref{Eq_4}, it is realized that this relation is valid in the case of infinite number of snapshots ($L=\infty$) in which the equations $E\{\gamma_{s_i} \mathbf{n}^H\} = 0$ and $E\{\mathbf{n} \mathbf{n}^H\} = \sigma_n^2 \mathbf{I}_N$ are hold. But in the practical case, where the number of snapshots is limited, an error term occurs in estimating the covariance matrix which is referred as subspace leakage in literature \cite{johnson2008music,shaghaghi2015subspace}. In this case, rewriting \eqref{Eq_4} gives
	\begin{equation}
		\label{Eq_28}
		\mathbf{R} = E[\mathbf{g}\mathbf{g}^H] = \sum_{i=1}^{k}\sigma_{s_i}^2 \boldsymbol{a}(s_{m_i})\boldsymbol{a}^H(s_{m_i}) + \sigma_n^2 \mathbf{I}_N + \Delta, 
	\end{equation}  
	where $\Delta$ is the estimation error term. In fact, by using the correlation subspace in estimating the covariance matrix, the error component is weakened, although it is not completely eliminated. It is useful to evaluate the performance of the covariance estimation methods regarding the involved parameters. To measure the efficiency of the covariance matrix estimation using the subspace-based methods, the subspace distance criterion is defined as follows \cite{jain2013low,rahmani2016subspace}
	\begin{equation}
		\label{Eq_31}
		dist(\mathbf{R},\mathbf{\hat{R}}) = \|\mathbf{E_s}^H\mathbf{\hat{E}_n}\|.
	\end{equation} 
	That $\mathbf{E_s}$ is the ideal matrix signal subspace that is derived from equation \eqref{Eq_5} and $\hat{\mathbf{E_n}}$ is the noise subspace of estimated $\mathbf{\hat{R}}$. The less the distance, the more the orthogonality of the two subspaces and a more accurate estimate of the covariance matrix is expected. In the optimal and suboptimal methods, the variance cancellation and positive semidefiniteness are two important factors that their impact should be investigated.
	\subsubsection{Noise variance cancellation}
	According to the role of the canceling noise variance term $- \sigma_n^2\mathbf{I}_N$ in the equations \eqref{Eq_10} and \eqref{Eq_13}, it should be carefully considered. Cancellation of the noise variance term is justifiable in the ideal case ($L=\infty$) but care must be taken to the removal of this term in the limited L values. To investigate this issue, the efficiency of the covariance estimator should be measured in the two cases of canceled and non-canceled noise variance. For the non-canceled noise variance case, the expressions of \eqref{Eq_10} and \eqref{Eq_17} are respectively as
	
	\begin{equation}
		\label{Eq_29}		
		\enspace \mathbf{R}_{\mathrm{P}}=\arg\min_{\mathbf{R}} \|\hat{\mathbf{R}} - \mathbf{R}\|_2^2,  
	\end{equation}
	
	\begin{equation}
		\label{Eq_30}
		\mathrm{vec} \left(\mathbf{R}_{\mathrm{Pa}}\right)=\mathbf{Q}\mathbf{Q}^{\dagger}\mathrm{vec} \left(\hat{\mathbf{R}} \right).
	\end{equation}	
	
\begin{figure}
		\centering
		\includegraphics[width=0.76\linewidth]{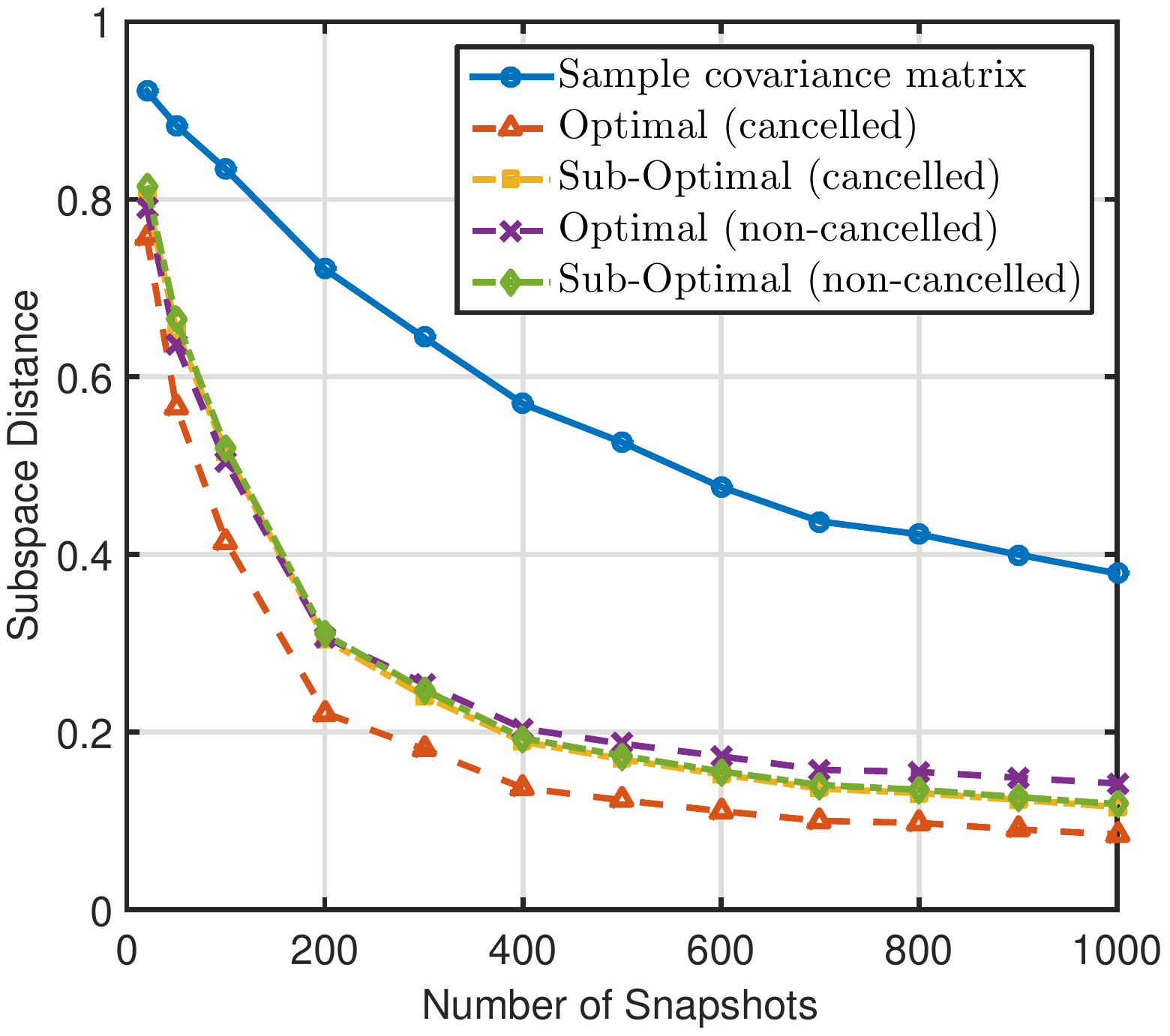}
		\caption{Subspace distance of different covariance matrix estimation methods regarding the noise variance cancellation under SNR = -6 dB.}
		\label{fig:subspacedistlowsnr}
	\end{figure}

	Fig. \ref{fig:subspacedistlowsnr} shows the results of the subspace distance at SNR = -6 dB. The estimation methods here are the sample covariance matrices and the optimal and suboptimal solutions of correlation subspace. In the optimization problem $\mathrm{(P)}$, the value of noise variance is important and hence, should be estimated. Noise variance is estimated using ML method as 
	\begin{equation}
		\label{Eq_30}
		\hat{\sigma}_n^2 = \frac{1}{N-k}\sum_{i=k+1}^{N}\lambda_i,
	\end{equation}
	where $\lambda_i$ are $N - k$ smallest eigenvalues of the sample covariance matrix.	
	
	To investigate the effect of canceling the noise variance, each of the two optimal and suboptimal solutions has been examined in two cases of canceled and non-canceled noise variance terms.
	
	As shown in Fig. \ref{fig:subspacedistlowsnr}, the subspace distance of the correlation subspace is less than the subspace distance of the sample covariance matrix and hence, higher efficiency is expected for the correlation subspace method.
	But in comparing the two variants of canceled or non-canceled of the noise variance in the correlation subspace, it is observed that the optimal solution, in which the noise variance term is subtracted from the sample covariance matrix, has the least subspace distance. But in the suboptimal solution, the two modes of canceled noise variance and non-canceled do not differ and their curves almost coincide. Also, the optimal method, in which the noise variance term is not subtracted from the covariance matrix, has the least efficiency. Therefore, it is concluded that for using the optimal solution, the noise variance should be canceled, but for the suboptimal one, the noise variance subtraction is not necessary.
	
	\subsubsection{Positive semidefinite constraint}
\begin{figure}
		\centering
		\includegraphics[width=0.76\linewidth]{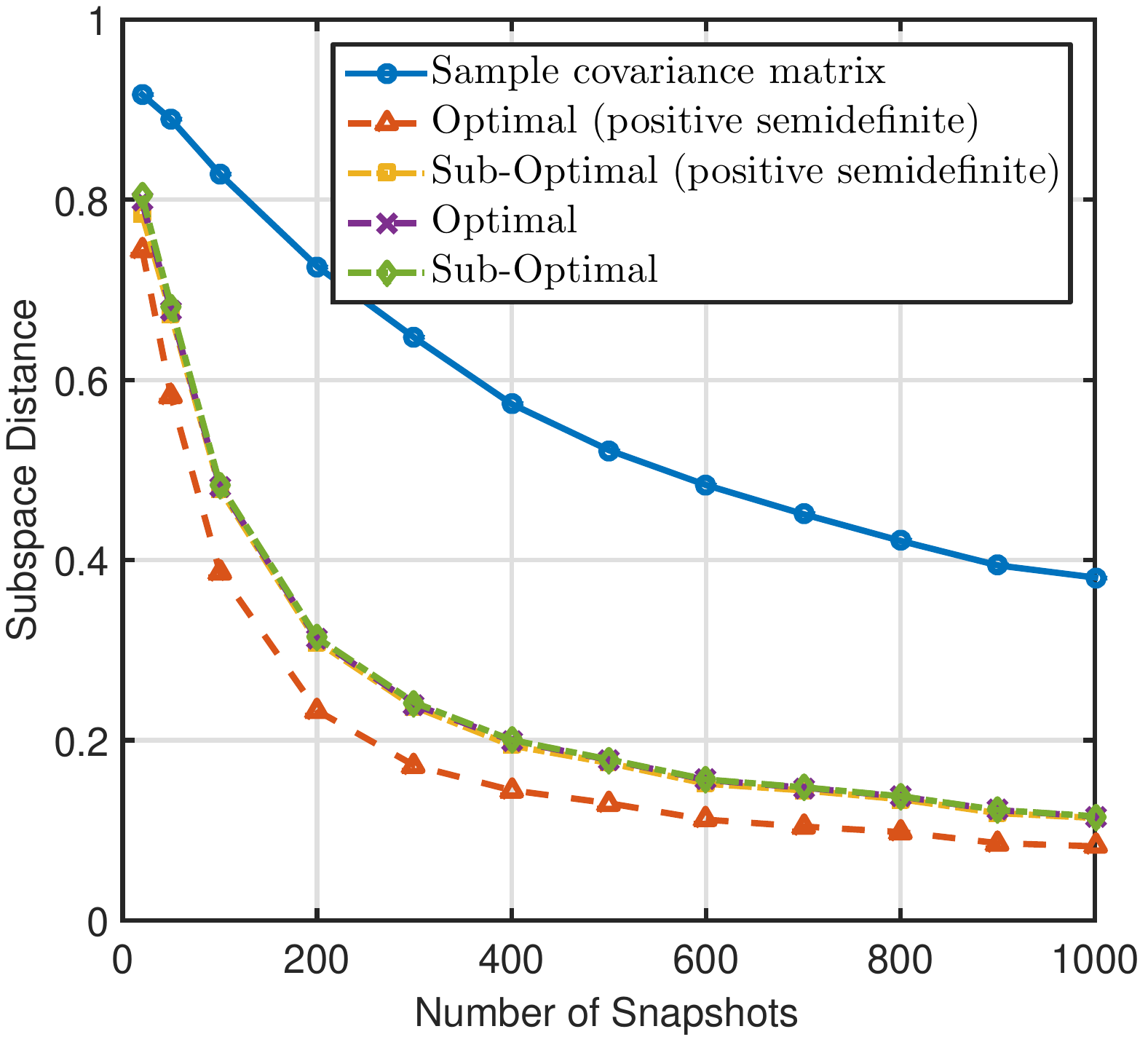}
		\caption{Subspace distance of different covariance matrix estimation methods regarding the positive semidefiniteness under SNR = -6 dB.}
		\label{fig:subspacedistlowsnrpd}
	\end{figure}

		In addition to the experiments carried out for investigating the role of canceling the noise variance, we also considered the impact of positive semidefiniteness on the performance of the subspace correlation method. Due to the constraints \eqref{Eq_12} and \eqref{Eq_16} in optimal and suboptimal methods, the desired ideal covariance matrix should be a positive semidefinite matrix. A question may be raised about the necessity of this condition and its effect on the performance of covariance estimation. To find an answer, we made use of subspace distance to measure the performance of variants of the correlation subspace method, i.e. the optimal and suboptimal solutions and applying or not applying the positive semidefiniteness constraint. The result is shown in Fig. \ref{fig:subspacedistlowsnrpd}. 
		
		As seen, the performance of the positive semidefinite optimal solution is the highest but, those of optimal solution without positive semidefiniteness and the two variants of suboptimal method are almost coincided and have a lower performance. Hence, in the optimal solution, imposing positive semidefiniteness in \eqref{Eq_12} makes the estimation result improve but, for the suboptimal method, this constraint is not necessary. However, due to the lack of necessity for the subtraction of noise variance in the suboptimal solution, it can be shown that the covariance matrix resulting from equation \eqref{Eq_30} is always positive semidefinite. A simple justification is that both $\mathbf{Q}\mathbf{Q}^{\dagger}$ and $\hat{\mathbf{R}}$ are positive semidefinite. Hence there is no need for an eigendecomposition and pick reconstruct a new positive semidefinite matrix. Accordingly, a simplified suboptimal solution of correlation subspace can be proposed regardless of the noise variance estimation and imposing positive semidefiniteness property as shown in Algorithm \ref{alg.CORRSUBSIMPLIFIED}.
	
	\begin{algorithm}[t!] 
		\caption{Covariance matrix estimation using simplified correlation subspace} 
		\label{alg.CORRSUBSIMPLIFIED} 
		\input{algs/CORRSUBSIMPLIFIED}
	\end{algorithm} 	
	
	\section{Conclusion}\label{sec6}	
	A new sequential MUSIC algorithm named RCC-MUSIC was introduced for the detection of point-like scatterers. In this study, we evaluate the performance of the proposed method in terms of the estimation accuracy of the location of targets. We have investigated the estimation accuracy considering the impact of SNR and the distance of targets. Compared with the other successive cancellation-based detectors, RCC-MUSIC has a lower estimation error due to two main reasons. First, the sequential MUSIC algorithms are advantageous over SOMP-based SGLRTC as the detection is applied over the MUSIC power spectrum rather than the beamforming spectrum. Second is that among the sequential MUSIC algorithms, RCC-MUSIC has the best quality which is the result of performing cancellation directly on the covariance matrix instead of the cancellation over the derived power spectrum.	
	
	Also, we have utilized the elegant method of correlation subspace for estimation of the covariance matrix to further increase the resolution. By the correlation subspace method, the initially estimated sample covariance matrix is fitted to the ideal form and consequently, a denoised covariance matrix is obtained. We considered the two solutions, optimal and suboptimal for the implementation of correlation subspace. It is observed that although the optimal solution is more efficient for the classical MUSIC methods, their performance is almost the same in the RCC-MUSIC method. Moreover, the performance of the covariance matrix estimation has been measured using the subspace distance criterion which helped to reach a simpler implementation of the suboptimal method. By combining the suboptimal correlation subspace with the proposed RCC-MUSIC method, a high-quality estimator is achieved. It is shown that although the estimation error of the proposed method is bounded in high SNR, it is a good choice for low and moderate SNR conditions with an acceptable computational cost.
	
	\appendix
	\section{Correlation subspace}
	\label{App1}
	By considering the expression of covariance matrix in \eqref{Eq_4} and applying vectorization, we have
	\begin{equation}
		\label{Eq_A1}
		\mathrm{vec} \left(\mathbf{R} - \sigma_n^2 \mathbf{I}_N \right) = \sum_{i=1}^{k}\sigma_{s_i}^2 \mathbf{c}(m_i)  
	\end{equation} 
	where $\mathbf{c}(m_i)$ is called the correlation vector and it is defined as \cite{rahmani2016subspace}
	\begin{equation}
		\label{Eq_A2}
		\mathbf{c}(m_i)  \triangleq \mathrm{vec} \left(\boldsymbol{a}(s_{m_i})\boldsymbol{a}^H(s_{m_i})\right)
	\end{equation} 
	and $\mathrm{vec()}$ is the vectorization operator. Equation \eqref{Eq_A1} indicates that $\mathrm{vec} \left(\mathbf{R} - \sigma_n^2 \mathbf{I}_N \right) $ is always spanned by correlation vectors, i.e.
	\begin{equation}
		\label{Eq_A3}
		\mathrm{vec} \left(\mathbf{R} - \sigma_n^2 \mathbf{I}_N \right) \in \mathrm{span}\{\mathbf{c}(m_i): i = 1, 2, \ldots ,k\}
	\end{equation} 
	
	On the other hand, the correlation vectors belong to a larger set called the correlation subspace, which is defined as  \cite{liu2017correlation}
	\begin{equation}
		\label{Eq_A4}
		\mathcal{CS} \triangleq \mathrm{span}\{\mathbf{c}(i): i = 1, 2, \ldots ,M\}
	\end{equation} 
	
	It is shown that the correlation subspace is obtained as the following \cite{rahmani2016subspace,liu2017correlation}
	\begin{equation}
		\label{Eq_A5}
		\mathcal{CS}=\mathrm{col}(\mathbf{B})
	\end{equation} 	
	where $\mathbf{B}$ is a positive semi-definite matrix defined as
	\begin{equation}
		\label{Eq_A6}
		\mathbf{B} = \sum_{i=1}^{M}\mathbf{c}(i)\mathbf{c}(i)^H
	\end{equation} 
	and $\mathrm{col}(\mathbf{B})$ is the column space of the matrix B. Equation \eqref{Eq_A5} states that the correlation subspace is spanned by the column vectors of matrix B. On the other hand, the ideal covariance matrix is a $N \times N$ Toeplitz matrix which has $2N-1$ distinct elements. Hence, the vectorized covariance matrix can be expressd by a $(2N-1)$-dimensional space (for further details, see \cite[Appendix, lemma 1]{rahmani2014robust}). As a result, the correlation subspace can be obtained by eigenvectors corresponding to $2N-1$ significant eigenvalues of $\mathbf{B}$. To this end, eigendecomposition of matrix $\mathbf{B}$ is performed
	\begin{equation}
		\label{Eq_A7}
		\mathbf{B} = 
		\left[\mathbf{Q} , \mathbf{Q}^\perp\right]
		\left[\def\arraystretch{0.6}
		\begin{array}{cc}
			\mathbf{\Lambda} & \mathbf{0} \\
			\mathbf{0} & \mathbf{0}
		\end{array}\right]
		\left[\mathbf{Q} , \mathbf{Q}^\perp\right]^H
	\end{equation}
	where $\mathbf{\Lambda}$ is the non-zero eigenvalues, $\mathbf{Q}$ is the corresponding eigenvectors and $\mathbf{Q}^\perp$ is the eigenvectors corresponding to the zero eigenvalues.	Consequently, $\mathbf{Q}$, which is a $N^2 \times (2N-1)$ matrix, includes basis vectors of the correlation subspace, i.e., $\mathcal{CS}=\mathrm{col}(\mathbf{Q})$.

	\bibliography{Library}
\end{document}

%% file: algs/CORRSUB.tex
\begin{algorithmic}[1]
	\STATE $\mathbf{Input}: \hat{\mathbf{R}},\mathbf{A},k$
	\STATE $[\mathbf{U},\mathbf{S}] = eig(\hat{\mathbf{R}})$ 
	\STATE $\hat{\sigma}_n^2 = \frac{1}{N-k}\sum_{i=k+1}^{N}\mathbf{S}_{ii} $
	\STATE $\mathbf{c}(i)  = \mathrm{vec} \left(\boldsymbol{a}(s_i)\boldsymbol{a}^H(s_i)\right),\hspace{0.15cm} i \in [1,M]$
	\STATE $\mathbf{B} = \sum_{i=1}^{M}\mathbf{c}(i)\mathbf{c}(i)^H$
	\STATE $[\mathbf{U},\mathbf{S}] = eig(\mathbf{B}), \mathbf{Q} = \mathbf{U}(:,1:2N-1)$
	\STATE $\mathrm{vec} \left(\mathbf{R}_{\mathrm{Pa}}\right)=\mathbf{Q}\mathbf{Q}^H\mathrm{vec} \left(\hat{\mathbf{R}} - \hat{\sigma}_n^2 \mathbf{I}_N \right)$
	\STATE $[\mathbf{U},\mathbf{S}] = eig(\mathbf{R}_{\mathrm{Pa}}),\mathrm{Ppi} = \{i|\mathbf{S}_{ii} \geq 0\}$		
	\STATE $\mathbf{R}_{\mathrm{Pb}}=\mathbf{U}(:,\mathrm{Ppi})\mathbf{S}(\mathrm{Ppi},\mathrm{Ppi})\mathbf{U}^H(:,\mathrm{Ppi})$
	\STATE $\mathbf{Output}: \mathbf{R}_{\mathrm{P}}=\mathbf{R}_{\mathrm{Pb}}$		
\end{algorithmic}

%% file: algs/RAPMUSIC.tex
	\begin{algorithmic}[1]
		\STATE $\mathbf{Input}: \mathbf{R},\mathbf{A},k$
		\STATE Initialization: $i=1,\Omega^{0}=\emptyset,\Pi_{\Omega^{0}}=\emptyset$
		\STATE $[\mathbf{U},\mathbf{S}] = eig(\mathbf{R}), \mathbf{U}_s =  \mathbf{U}(:,1:k), \mathbf{U}_n =  \mathbf{U}(:,k+1:N)$
		\WHILE{$ \hspace{0.15cm} (i \leq k \hspace{0.15cm}) $}
		\STATE $\Pi_{\Omega^{i-1}} = \mathbf{A}_{\Omega^{i-1}}[\mathbf{A}^H_{\Omega^{i-1}}\mathbf{A}_{\Omega^{i-1}}]^{-1}\mathbf{A}^H_{\Omega^{i-1}}$
		\STATE $m_i = \arg\max_n\{\boldsymbol{a}(s_n)^H(\mathbf{I}_N - \Pi_{\Omega^{i-1}})\mathbf{U}_s\mathbf{U}_s^H(\mathbf{I}_N - \Pi_{\Omega^{i-1}})\boldsymbol{a}(s_n)\}$
		\STATE $\Omega^{i} = \Omega^{i-1} \bigcup m_i$
		\STATE $i=i+1$
		\ENDWHILE	
		\STATE‌ $\mathbf{Output}:\Omega^{k}=\{m_1,m_2,\ldots,m_k\}$		
	\end{algorithmic}

%% file: algs/RCCMUSIC.tex
	\begin{algorithmic}[1]
		\STATE $\mathbf{Input}: \mathbf{g}, \mathbf{R},\mathbf{A}, k$
		\STATE Initialization: $i=1,\Omega^{0}=\emptyset$
		\WHILE{$ \hspace{0.15cm} (i \leq k \hspace{0.15cm}) $}
		\STATE $\Lambda = \frac{1}{L}\sum_{l=1}^{L}|(\mathbf{A}^H_{\Omega^{(i-1)}}\mathbf{A}_{\Omega^{(i-1)}})^{-1}\mathbf{A}^H_{\Omega^{(i-1)}}\mathbf{g}(l)|^2$
		\STATE $\mathbf{R^{(i)}} = \mathbf{R}$
		\FOR{$p=1; p:=p+1; p < i$}
		\STATE $\mathbf{R^{(i)}} = \mathbf{R^{(i)}} - \Lambda(p)\boldsymbol{a}(s_{m_p})\boldsymbol{a}^H(s_{m_p})$
		\ENDFOR
		\STATE  $[\mathbf{U},\mathbf{S}] = eig(\mathbf{R^{(i)}}), \mathbf{U}_s =  \mathbf{U}(:,1:k-i+1), \mathbf{U}_n = \mathbf{U}(:,k-i+2:N)$
		\STATE $m_i = arg\max_n\{\boldsymbol{a}(s_n)^H\mathbf{U}_s\mathbf{U}_s^H\boldsymbol{a}(s_n)\}$
		\STATE $\Omega^{i} = \Omega^{i-1} \bigcup m_i$
		\STATE $i=i+1$
		\ENDWHILE	
		\STATE‌ $\mathbf{Output}:\Omega^{k}=\{m_1,m_2,\ldots,m_k\}$			
	\end{algorithmic}

%% file: algs/CORRSUBSIMPLIFIED.tex
\begin{algorithmic}[1]
	\STATE $\mathbf{Input}: \hat{\mathbf{R}},\mathbf{A},k$
	\STATE $\mathbf{c}(i)  = \mathrm{vec} \left(\boldsymbol{a}(s_i)\boldsymbol{a}^H(s_i)\right),\hspace{0.15cm} i \in [1,M]$
	\STATE $\mathbf{B} = \sum_{i=1}^{M}\mathbf{c}(i)\mathbf{c}(i)^H$
	\STATE $[\mathbf{U},\mathbf{S}] = eig(\mathbf{B}), \mathbf{Q} = \mathbf{U}(:,1:2N-1)$
	\STATE $\mathrm{vec} \left(\mathbf{R}_{\mathrm{P}}\right)=\mathbf{Q}\mathbf{Q}^H\mathrm{vec} \left(\hat{\mathbf{R}} \right)$
	\STATE $\mathbf{Output}: \mathbf{R}_{\mathrm{P}}$		
\end{algorithmic}

%% file: Digital_Signal_Processing.bbl
\begin{thebibliography}{10}
\expandafter\ifx\csname url\endcsname\relax
  \def\url#1{\texttt{#1}}\fi
\expandafter\ifx\csname urlprefix\endcsname\relax\def\urlprefix{URL }\fi
\expandafter\ifx\csname href\endcsname\relax
  \def\href#1#2{#2} \def\path#1{#1}\fi

\bibitem{moreira2013tutorial}
A.~Moreira, P.~Prats-Iraola, M.~Younis, G.~Krieger, I.~Hajnsek, K.~P.
  Papathanassiou, A tutorial on synthetic aperture radar, IEEE Geoscience and
  remote sensing magazine 1~(1) (2013) 6--43.

\bibitem{reigber2000first}
A.~Reigber, A.~Moreira, First demonstration of airborne sar tomography using
  multibaseline l-band data, IEEE Transactions on Geoscience and Remote Sensing
  38~(5) (2000) 2142--2152.

\bibitem{fornaro2005three}
G.~Fornaro, F.~Lombardini, F.~Serafino, Three-dimensional multipass sar
  focusing: Experiments with long-term spaceborne data, IEEE Transactions on
  Geoscience and Remote Sensing 43~(4) (2005) 702--714.

\bibitem{fornaro2014tomographic}
G.~Fornaro, F.~Lombardini, A.~Pauciullo, D.~Reale, F.~Viviani, Tomographic
  processing of interferometric sar data: Developments, applications, and
  future research perspectives, IEEE Signal Processing Magazine 31~(4) (2014)
  41--50.

\bibitem{siddique2016single}
M.~A. Siddique, U.~Wegm{\"u}ller, I.~Hajnsek, O.~Frey, Single-look sar
  tomography as an add-on to psi for improved deformation analysis in urban
  areas, IEEE Transactions on Geoscience and Remote Sensing 54~(10) (2016)
  6119--6137.

\bibitem{fornaro2003three}
G.~Fornaro, F.~Serafino, F.~Soldovieri, Three-dimensional focusing with
  multipass sar data, IEEE Transactions on Geoscience and Remote Sensing 41~(3)
  (2003) 507--517.

\bibitem{lombardini2003adaptive}
F.~Lombardini, A.~Reigber, Adaptive spectral estimation for multibaseline sar
  tomography with airborne l-band data, in: Geoscience and Remote Sensing
  Symposium, 2003. IGARSS'03. Proceedings. 2003 IEEE International, Vol.~3,
  IEEE, 2003, pp. 2014--2016.

\bibitem{guillaso2005polarimetric}
S.~Guillaso, A.~Reigber, Polarimetric sar tomography, in: in Proc. of POLINSAR,
  Frascati, Italy, 2005.

\bibitem{budillon2016glrt}
A.~Budillon, G.~Schirinzi, Glrt based on support estimation for multiple
  scatterers detection in sar tomography, IEEE Journal of Selected Topics in
  Applied Earth Observations and Remote Sensing 9~(3) (2016) 1086--1094.

\bibitem{pauciullo2012detection}
A.~Pauciullo, D.~Reale, A.~De~Maio, G.~Fornaro, Detection of double scatterers
  in sar tomography, IEEE Transactions on Geoscience and Remote Sensing 50~(9)
  (2012) 3567--3586.

\bibitem{zhu2010very}
X.~X. Zhu, R.~Bamler, Very high resolution spaceborne sar tomography in urban
  environment, IEEE Transactions on Geoscience and Remote Sensing 48~(12)
  (2010) 4296--4308.

\bibitem{zhu2012super}
X.~X. Zhu, R.~Bamler, Super-resolution power and robustness of compressive
  sensing for spectral estimation with application to spaceborne tomographic
  sar, IEEE Transactions on Geoscience and Remote Sensing 50~(1) (2012)
  247--258.

\bibitem{pauciullo2018multi}
A.~Pauciullo, D.~Reale, W.~Franz{\'e}, G.~Fornaro, Multi-look in glrt-based
  detection of single and double persistent scatterers, IEEE Transactions on
  Geoscience and Remote Sensing 56~(9) (2018) 5125--5137.

\bibitem{schmide1979multiple}
R.~Schmide, Multiple emitter location and signal parameter estimation-radc
  spectrum estimation workshop, IEEE Trans. Antennas Propag 34 (1979) 276--280.

\bibitem{stoica1989music}
P.~Stoica, A.~Nehorai, Music, maximum likelihood, and cramer-rao bound, IEEE
  Transactions on Acoustics, speech, and signal processing 37~(5) (1989)
  720--741.

\bibitem{krim1996two}
H.~Krim, M.~Viberg, Two decades of array signal processing research: the
  parametric approach, IEEE signal processing magazine 13~(4) (1996) 67--94.

\bibitem{kundu1996modified}
D.~Kundu, Modified music algorithm for estimating doa of signals, Signal
  processing 48~(1) (1996) 85--90.

\bibitem{zhang2018localization}
X.~Zhang, W.~Chen, W.~Zheng, Z.~Xia, Y.~Wang, Localization of near-field
  sources: A reduced-dimension music algorithm, IEEE Communications Letters
  22~(7) (2018) 1422--1425.

\bibitem{asghari2022ecf}
M.~Asghari, M.~Zareinejad, S.~M. Rezaei, H.~Amindavar, Ecf-music: An empirical
  characteristic function based direction of arrival (doa) estimation in the
  presence of impulsive noise, Digital Signal Processing 123 (2022) 103440.

\bibitem{lombardini2003reflectivity}
F.~Lombardini, M.~Montanari, F.~Gini, Reflectivity estimation for multibaseline
  interferometric radar imaging of layover extended sources, IEEE Transactions
  on Signal Processing 51~(6) (2003) 1508--1519.

\bibitem{tropp2006algorithms}
J.~A. Tropp, A.~C. Gilbert, M.~J. Strauss, Algorithms for simultaneous sparse
  approximation. part i: Greedy pursuit, Signal processing 86~(3) (2006)
  572--588.

\bibitem{dai2009subspace}
W.~Dai, O.~Milenkovic, Subspace pursuit for compressive sensing signal
  reconstruction, IEEE transactions on Information Theory 55~(5) (2009)
  2230--2249.

\bibitem{oh1993sequential}
S.~K. Oh, C.~K. Un, A sequential estimation approach for performance
  improvement of eigenstructure-based methods in array processing, IEEE
  Transactions on Signal Processing 41~(1) (1993) 457.

\bibitem{stoica1995improved}
P.~Stoica, P.~Handel, A.~Nehorai, Improved sequential music, IEEE Transactions
  on Aerospace and Electronic Systems 31~(4) (1995) 1230--1239.

\bibitem{mosher1999source}
J.~C. Mosher, R.~M. Leahy, Source localization using recursively applied and
  projected (rap) music, IEEE Transactions on signal processing 47~(2) (1999)
  332--340.

\bibitem{davies2012rank}
M.~E. Davies, Y.~C. Eldar, Rank awareness in joint sparse recovery, IEEE
  Transactions on Information Theory 58~(2) (2012) 1135--1146.

\bibitem{dehghani2018fomp}
M.~Dehghani, K.~Aghababaiyan, Fomp algorithm for direction of arrival
  estimation, Physical Communication 26 (2018) 170--174.

\bibitem{naghavi2020super}
A.~Naghavi, M.~S. Fazel, M.~Beheshti, E.~Yazdian, Super-resolving multiple
  scatterers detection in synthetic aperture radar tomography assisted by
  correlation information, Journal of Applied Remote Sensing 14~(3) (2020)
  034517.

\bibitem{zhao2021direction}


\bibitem{shaghaghi2015subspace}
M.~Shaghaghi, S.~A. Vorobyov, Subspace leakage analysis and improved doa
  estimation with small sample size, IEEE Transactions on Signal Processing
  63~(12) (2015) 3251--3265.

\bibitem{yazdian2012source}
E.~Yazdian, S.~Gazor, H.~Bastani, Source enumeration in large arrays using
  moments of eigenvalues and relatively few samples, IET signal processing
  6~(7) (2012) 689--696.

\bibitem{wikes1988iterated}
D.~Wikes, M.~H. Hayes, Iterated toeplitz approximation of covariance matrices,
  in: ICASSP-88., International Conference on Acoustics, Speech, and Signal
  Processing, IEEE, 1988, pp. 1663--1666.

\bibitem{li1999computationally}
H.~Li, P.~Stoica, J.~Li, Computationally efficient maximum likelihood
  estimation of structured covariance matrices, IEEE Transactions on Signal
  Processing 47~(5) (1999) 1314--1323.

\bibitem{chachlakis2021structured}
D.~G. Chachlakis, P.~P. Markopoulos, Structured autocorrelation matrix
  estimation for coprime arrays, Signal Processing 183 (2021) 107987.

\bibitem{wang2021fda}
L.~Wang, W.-Q. Wang, Y.~Zhou, Fda-mimo radar covariance matrix estimation via
  shrinkage processing, Digital Signal Processing 118 (2021) 103206.

\bibitem{lounici2014high}
K.~Lounici, High-dimensional covariance matrix estimation with missing
  observations, Bernoulli 20~(3) (2014) 1029--1058.

\bibitem{chu2003structured}
M.~T. Chu, R.~E. Funderlic, R.~J. Plemmons, Structured low rank approximation,
  Linear algebra and its applications 366 (2003) 157--172.

\bibitem{liu2021rank}
S.~Liu, Z.~Mao, Y.~D. Zhang, Y.~Huang, Rank minimization-based toeplitz
  reconstruction for doa estimation using coprime array, IEEE Communications
  Letters 25~(7) (2021) 2265--2269.

\bibitem{hippert2022robust}
A.~Hippert-Ferrer, M.~N. El~Korso, A.~Breloy, G.~Ginolhac, Robust low-rank
  covariance matrix estimation with a general pattern of missing values, Signal
  Processing (2022) 108460.

\bibitem{bien2011sparse}
J.~Bien, R.~J. Tibshirani, Sparse estimation of a covariance matrix, Biometrika
  98~(4) (2011) 807--820.

\bibitem{romero2015compressive}
D.~Romero, D.~D. Ariananda, Z.~Tian, G.~Leus, Compressive covariance sensing:
  Structure-based compressive sensing beyond sparsity, IEEE signal processing
  magazine 33~(1) (2015) 78--93.

\bibitem{prasanna2021mmwave}
D.~Prasanna, C.~R. Murthy, mmwave channel estimation via compressive covariance
  estimation: Role of sparsity and intra-vector correlation, IEEE Transactions
  on Signal Processing 69 (2021) 2356--2370.

\bibitem{burg1982estimation}
J.~P. Burg, D.~G. Luenberger, D.~L. Wenger, Estimation of structured covariance
  matrices, Proceedings of the IEEE 70~(9) (1982) 963--974.

\bibitem{jansson2000structured}
M.~Jansson, B.~Ottersten, Structured covariance matrix estimation: A parametric
  approach, in: 2000 IEEE International Conference on Acoustics, Speech, and
  Signal Processing. Proceedings (Cat. No. 00CH37100), Vol.~5, IEEE, 2000, pp.
  3172--3175.

\bibitem{rahmani2016subspace}
M.~Rahmani, G.~K. Atia, A subspace method for array covariance matrix
  estimation, in: 2016 IEEE Sensor Array and Multichannel Signal Processing
  Workshop (SAM), IEEE, 2016, pp. 1--5.

\bibitem{liu2017correlation}
C.-L. Liu, P.~Vaidyanathan, Correlation subspaces: Generalizations and
  connection to difference coarrays, IEEE Transactions on Signal Processing
  65~(19) (2017) 5006--5020.

\bibitem{wax1985detection}
M.~Wax, T.~Kailath, Detection of signals by information theoretic criteria,
  IEEE Transactions on Acoustics, Speech, and Signal Processing 33~(2) (1985)
  387--392.

\bibitem{kay1993fundamentals}
S.~M. Kay, Fundamentals of statistical signal processing: Estimation theory,
  ptr prentice hall, Inc., Englewood Cliffs, NJ (1993).

\bibitem{li1996efficient}
J.~Li, P.~Stoica, Efficient mixed-spectrum estimation with applications to
  target feature extraction, IEEE transactions on signal processing 44~(2)
  (1996) 281--295.

\bibitem{gini2002layover}
F.~Gini, F.~Lombardini, M.~Montanari, Layover solution in multibaseline sar
  interferometry, IEEE Transactions on Aerospace and Electronic Systems 38~(4)
  (2002) 1344--1356.

\bibitem{schmidt1986multiple}
R.~Schmidt, Multiple emitter location and signal parameter estimation, IEEE
  transactions on antennas and propagation 34~(3) (1986) 276--280.

\bibitem{akaike1974new}
H.~Akaike, A new look at the statistical model identification, in: Selected
  Papers of Hirotugu Akaike, Springer, 1974, pp. 215--222.

\bibitem{rissanen1978modeling}
J.~Rissanen, Modeling by shortest data description, Automatica 14~(5) (1978)
  465--471.

\bibitem{meyer2000matrix}
C.~D. Meyer, Matrix analysis and applied linear algebra, Vol.~71, Siam, 2000.

\bibitem{abdi2007method}
H.~Abdi, et~al., The method of least squares, Encyclopedia of Measurement and
  Statistics. CA, USA: Thousand Oaks (2007).

\bibitem{stoica2010spice}
P.~Stoica, P.~Babu, J.~Li, Spice: A sparse covariance-based estimation method
  for array processing, IEEE Transactions on Signal Processing 59~(2) (2010)
  629--638.

\bibitem{jatoi2014survey}
M.~A. Jatoi, N.~Kamel, A.~S. Malik, I.~Faye, T.~Begum, A survey of methods used
  for source localization using eeg signals, Biomedical Signal Processing and
  Control 11 (2014) 42--52.

\bibitem{golub2003separable}
G.~Golub, V.~Pereyra, Separable nonlinear least squares: the variable
  projection method and its applications, Inverse problems 19~(2) (2003) R1.

\bibitem{rife1974single}
D.~Rife, R.~Boorstyn, Single tone parameter estimation from discrete-time
  observations, IEEE Transactions on information theory 20~(5) (1974) 591--598.

\bibitem{lee1992cramer}
H.~B. Lee, The cram{\'e}r-rao bound on frequency estimates of signals closely
  spaced in frequency, IEEE Transactions on Signal Processing 40~(6) (1992)
  1507--1517.

\bibitem{swingler1993frequency}
D.~N. Swingler, Frequency estimation for closely spaced sinsoids: Simple
  approximations to the cram{\'e}r-rao lower bound, IEEE transactions on signal
  processing 41~(1) (1993) 489.

\bibitem{johnson2008music}
B.~A. Johnson, Y.~I. Abramovich, X.~Mestre, Music, g-music, and
  maximum-likelihood performance breakdown, IEEE Transactions on Signal
  Processing 56~(8) (2008) 3944--3958.

\bibitem{jain2013low}
P.~Jain, P.~Netrapalli, S.~Sanghavi, Low-rank matrix completion using
  alternating minimization, in: Proceedings of the forty-fifth annual ACM
  symposium on Theory of computing, 2013, pp. 665--674.

\bibitem{rahmani2014robust}
M.~Rahmani, M.~H. Bastani, Robust and rapid converging adaptive beamforming via
  a subspace method for the signal-plus-interferences covariance matrix
  estimation, IET Signal Processing 8~(5) (2014) 507--520.

\end{thebibliography}
